\documentclass[%
 reprint,
superscriptaddress,
 amsmath,amssymb,
 aps,
pra,
]{revtex4-2}

\usepackage[utf8]{inputenc}
\usepackage[T1]{fontenc}

\usepackage{graphicx}
\usepackage[caption=false]{subfig}  
\usepackage{color}
\usepackage{colortbl}
\usepackage{xcolor}
\usepackage{soul}
\usepackage{natbib}
\usepackage{amsmath}
\usepackage{mathtools}
\usepackage{bbm}
\usepackage{physics}
\usepackage[normalem]{ulem}
\usepackage[ruled,vlined]{algorithm2e}

\usepackage[caption=false]{subfig}
\usepackage[hypertexnames=false]{hyperref}

\usepackage{enumitem}  
\usepackage{cleveref}
\usepackage{braket}

\crefname{figure}{Fig.}{Figures}
\crefname{equation}{Eq.}{Equations}
\crefname{section}{Section}{Sections}
\crefname{appendix}{Appendix}{Appendices}

\newcommand{\tfinal}{t_\textnormal{f}}
\newcommand{\im}{i}

\begin{document}
\title{Universal pulses for superconducting qudit ladder gates}

\author{Boxi Li}
\email{b.li@fz-juelich.de}
\affiliation{Forschungszentrum Jülich, Institute of Quantum Control (PGI-8), D-52425 Jülich, Germany}
\affiliation{Institute for Theoretical Physics, University of Cologne, D-50937 Cologne, Germany}
\author{F. A. Cárdenas-López}
\affiliation{Forschungszentrum Jülich, Institute of Quantum Control (PGI-8), D-52425 Jülich, Germany}
\author{Adrian Lupascu}
\affiliation{Institute for Quantum Computing, Department of Physics and Astronomy, and Waterloo Institute for Nanotechnology, University of Waterloo, Waterloo, Ontario, Canada N2L 3G1}
\author{Felix Motzoi}
\email{f.motzoi@fz-juelich.de}
\affiliation{Forschungszentrum Jülich, Institute of Quantum Control (PGI-8), D-52425 Jülich, Germany}
\affiliation{Institute for Theoretical Physics, University of Cologne, D-50937 Cologne, Germany}

\begin{abstract}
Qudits, generalizations of qubits to multi-level quantum systems, offer enhanced computational efficiency by encoding more information per lattice cell, avoiding costly swap operations and providing even exponential speedup in some cases. Utilizing the $d$-level manifold, however, requires high-speed gate operations because of the stronger decoherence at higher levels.
While analytical control methods have proven effective for qubits in achieving fast gates with minimal control errors, their extension to qudits is nontrivial due to the increased complexity of the energy level structure arising from additional ancillary states.
In this work, we present a universal pulse construction for generating rapid, high-fidelity unitary rotations between adjacent qudit levels, thereby providing a prescription for any gate in $SU(d)$.
Control errors in these operations are effectively analyzed within a four-level subspace, including two leakage levels with approximately opposite detuning.
By identifying the optimal degrees of freedom, we derive concise analytical pulse schemes that suppress multiple control errors and outperform existing methods.
Remarkably, our approach achieves consistent coherent error scaling across all levels, approaching the quantum speed limit independently of parameter variations between levels.
Numerical validation on transmon circuits demonstrates significant improvements in gate fidelity for various qudit sizes aiming for $10^{-4}$ error.
This method provides a scalable solution for improving qudit control and can be broadly applied to other quantum systems with ladder structures or operations involving multiple ancillary levels.
\end{abstract}

\maketitle

\section{Introduction}
Quantum computation and quantum information processing protocols often rely on qubits, or two-level systems, as the fundamental units of computation due to their simplicity and close analogy to classical computing. However, most quantum systems comprise more than just two levels. These additional quantum levels can also be used as an information register, which is known as a qudit, a generalization of the qubit to a $d$-level system. 
Exploring the full Hilbert space of qudits enables more efficient computation by increasing the amount of information stored per quantum unit.

Qudit-based quantum computation offers several known advantages over qubit-based approaches. For instance, a $d$-level system can encode $\log_2(d)$ qubits~\cite{Gottesman1999FaultTolerant}.
This has been exploited for efficient compilation of arbitrary unitaries, requiring an exponentially reduced number of circuit layers~~\cite{Di2015Optimal,Motzoi2017Linear,Cao2024Emulating,Galda2021Implementing,Lanyon2009Simplifying}, to simulate bosonic modes for studying light-matter processes~\cite{PhysRevLett.125.260511,PhysRevResearch.3.043212} and lattice gauge theories~\cite{RICO2018466, PhysRevResearch.3.043209, Meth2025Simulating}, for enhancing the robustness in quantum cryptography~\cite{PhysRevLett.88.127901,PhysRevLett.85.3313}, and for simplified implementation of quantum error correction protocols~\cite{Grace2006Encoding,Chiesa2020Molecular,Campbell2014Enhanced}. In general, the larger density of registers means that the connectivity of qubit-based architectures is increased, since neighboring qudits can share up to $d^2$ level couplings.  Qudit processors have been implemented across various physical platforms, including trapped-ions~\cite{Low2020Practical,Ringbauer2022Universal,Hrmo2023Native,Low2023Control}, Rydberg atoms~\cite{Gonzalez-Cuadra2022Hardware}, ultracold atomic mixtures~\cite{Hussain2018}, molecular spins~\cite{D2CP01228F,Biard2021Increasing}, photonic systems~\cite{Kues2017Onchip, Erhard2018Experimental,Luo2019Quantum,Davis2019PhotonMediated,Chi2022Programmable} and superconducting circuits~\cite{Blok2021Quantum,Liu2023Performing,Champion2025Efficient,Morvan2021Qutrit,Yurtalan2020Implementation,Kononenko2021Characterization,Yurtalan2021Characterization,Luo2023Experimental}; such implementations contribute to significant progress on qudit-based quantum computation.

Despite these advancements, maintaining coherent control of all the qudit levels poses complex new challenges.
In transmon superconducting circuits, where the quantum system is represented by a non-linear oscillator~\cite{Koch2007Chargeinsensitive}, each qudit operation needs to be addressed differently due to the varying surrounding level structure. Compared to a qubit operation,  the presence of additional leakage channels significantly, limits the gate performance, as shown in \cref{fig:qudit energy levels}a-c. For instance, it has been reported that the gate time of a single-qutrit gate is around 30~ns~\cite{Kononenko2021Characterization,Morvan2021Qutrit,Liu2023Performing}, which is three times longer than that required for single-qubit gates with state-of-the-art quantum control techniques~\cite{Chen2016Measuring}.
Therefore, developing quantum control protocols for qudits is crucial for making qudit computation practical. Of particular relevance is the Derivative Removal by Adiabatic Gate (DRAG) method~\cite{Motzoi2009Simple,Gambetta2011Analytic,Motzoi2013Improving,Theis2018Counteracting}, successfully employed in superconducting qubit systems to reduce leakage and phase errors.
DRAG's simplicity and flexibility allow engineering efficient pulses with easy-to-calibrate parameters, making it ubiquitous in the superconducting qubits platform~\cite{Chow2010Optimized,Lucero2010Reduced,DiCarlo2009Demonstration,Chen2016Measuring, Wei2022Hamiltonian}. The same advantages remain even with the presence of multiple error sources, whereby multiple DRAG corrections can be combined, offering efficient yet compact solutions~\cite{Motzoi2013Improving,Li2024Experimental}.

In this article, we investigate the energy structure and dominant error mechanisms of a transmon ladder-type qudit, a weakly anharmonic oscillator, under nearest-level driving.
We examine the relationship between the number of usable levels in a transmon qudit and its design parameters, which constrain coherence times and gate design.
We demonstrate that transmon qudit control can be studied within a four-level subspace involving two nearest-neighbour interactions and effectively reduced to an enhanced error model in an $SU(2)$ subspace.
Through a systematic study across various qudit sizes in transmon circuits, we find that porting the widely-used single-derivative DRAG method, as previously suggested in \cite{Gambetta2011Analytic} and experimentally implemented in~\cite{Champion2025Efficient, Wang2025High}, does not significantly improve fidelity in the qudit case because it is overconstrained by multiple leakage channels.

To overcome this limitation, we apply the DRAG framework to engineer level-independent and high-precision analytical control schemes for qudit systems within a driven ladder structure.
We adopt a recursive DRAG approach introduced in Ref.~\cite{Li2024Experimental} that incorporates higher-order derivatives, providing new degrees of freedom that are used to suppress both single- and multi-photon errors.

Remarkably, despite the presence of multiple parameters in the circuit description, we observe a universal behaviour in the error budget and pulse-specific quantum speed limits, i.e., the minimum gate duration required to achieve a target fidelity.
In particular, for a given drive protocol, the speed limit is found to be inversely proportional to the system nonlinearity, and it remains independent of other circuit parameters or the specific qudit transition targeted.
However, it is highly sensitive to the drive protocol, particularly to whether certain leakage channels are effectively suppressed.
We observe significant improvements in gate performance and successfully reduce gate times to mitigate dephasing caused by voltage fluctuations during gate implementation.
These results are broadly applicable to any qudit platform with multiple connected ancillary levels.

In the following, we start with the transmon model and derive the four-level effective Hamiltonian in \cref{sec:qudit model}.
Next, we introduce and explore the recursive DRAG method in detail and perform a systematic study of its performance in \cref{sec:recursive DRAG qudit}.
Significant improvement in fidelity is observed across a wide range of parameters, with a universal behaviour across all levels independent of parameter variations between levels.
In \cref{sec:other error}, we discuss other potential control errors beyond the two-level transition and provide a summary of our findings in \cref{sec:qudit conclusion}.

\section{Qudit Model for universal quantum gates}
\label{sec:qudit model}
We consider a superconducting transmon qudit, modeled as a weakly anharmonic oscillator with dominant nearest-level interactions, as illustrated in~\cref{fig:qudit energy levels}a. In the rotating frame, the nonlinearity of the system allows for the isolation of effective subspaces that capture the dynamics of specific transitions (\cref{fig:qudit energy levels}b and c).
However, residual couplings to non-computational states introduce additional error channels.
In this section, we derive an effective Hamiltonian for such systems, focusing on selective control of adjacent-level transitions and the constraints imposed by the energy structure and decoherence.

\subsection{Native gate set for superconducting qudit}

Our objective is that, for a transmon system, each individual ladder transition between adjacent states, $\ket{k}$ and $\ket{k+1}$, can be selectively controlled. This allows arbitrary unitaries on $SU(d)$ to be implemented.
Since a calibrated $\pi/2$ gate combined with a virtual Z gate~\cite{McKay2017Efficient} is a complete native gate set, as we show in \cref{sec:qudit decompostion}, our primary focus in the following study is on the $\pi/2$ gate. In addition, we also present results for the $\pi$ gate, which represents the greatest challenge among Givens rotations at fixed gate duration, as it requires the strongest drive amplitude and typically incurs the highest leakage rate. Overall, these methods can be extended to rotations of arbitrary angles, which are very helpful for reducing circuit compilation depths \cite{Preti2022Continuous}.

\subsection{The transmon Hamiltonian}
\begin{figure}[t]
    \centering
    \includegraphics[width=\linewidth]{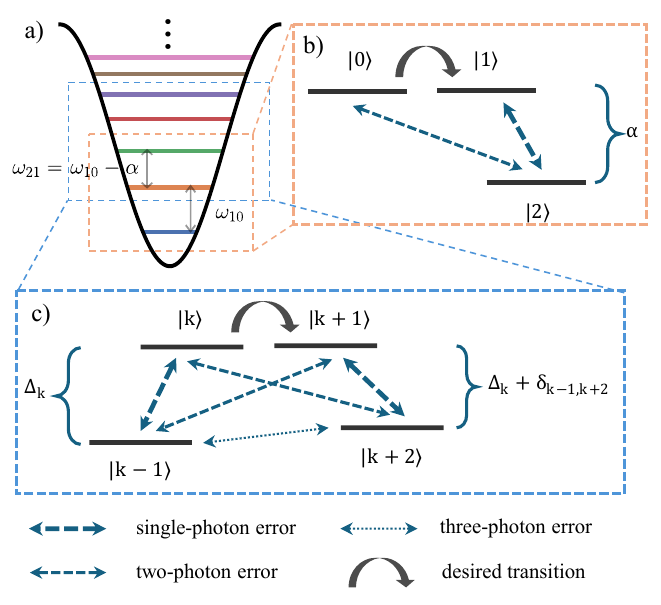}
    \includegraphics[width=\linewidth]{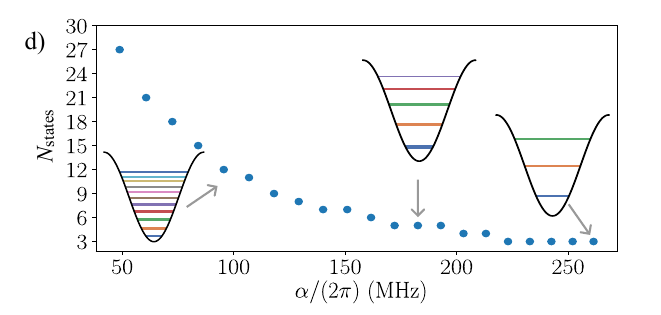}
    \caption{Energy structure of driving a two-level transition in a qudit system. (a) Typical energy structure of a transmon system.
    Energy levels and error transitions for (b) the ground and first excited states, and (c) general ladder transition between higher levels in the rotating frame.
    (d) The number of levels that can be used as a qudit quantum register as a function of the anharmonicity. The upper bound is set by the decoherence introduced by charge fluctuations. The detailed discussion can be found in ~\cref{sec:decoherence}.
    }
    \label{fig:qudit energy levels}
    \label{fig:Numberstates}
\end{figure}

We start by deriving the effective Hamiltonian for selectively driving the $\ket{k} \leftrightarrow \ket{k+1}$ transition in a superconducting transmon.  
A transmon nonlinear oscillator is described by the following Hamiltonian~\cite{Koch2007Chargeinsensitive}
\begin{equation}
    \hat{H} = 4E_{C}[\hat{n}-n_g(t)]^2-E_{J}\cos(\hat{\varphi})
\label{eq:transmon hamiltonian exact}
\end{equation}
where $E_{C}$ and $E_{J}$ represent the charge and Josephson energies, respectively, and $n_g(t)$ is the dimensionless gate voltage. The operator $\hat{n}$ is the charge operator, indicating the number of Cooper pairs on the island, and $\hat{\varphi}$ denotes the phase operator. For implementing single-qudit operations, we capacitively drive the transmon using $n_g(t)=n_{0}(t)\cos(\omega_{d}t)$ resulting in the Hamiltonian
\begin{eqnarray}
    \hat{H}_{{\rm{ctrl}}}=\Omega(t)\cos(\omega_{d}t)\hat{n}
\end{eqnarray}
where $\Omega(t)=-8E_Cn_{0}(t)$ is the complex drive envelope, and $\omega_{d}$ is the drive frequency.

When only the lowest few energy levels are considered, the transmon can be modelled as an approximate Duffing oscillator~\cite{Khani2009Optimal,Blais2021Circuit}. In this model, the control operator is expressed as $\hat{n}\propto(\hat{b}^{\dag}+\hat{b})$, where $\hat{b}$ is the annihilation operator of a linear oscillator. Consequently, the control Hamiltonian adopts a ladder configuration, connecting states $\ket{k}\leftrightarrow \ket{k\pm1}$.
The transmon Hamiltonian without external drive then simplifies to 
\begin{equation}
    \hat{H}_0^{\textnormal{duf}} = \omega_q \hat{b}^{\dagger}\hat{b}+\frac{\alpha}{2}\hat{b}^{\dagger}\hat{b}^{\dagger}\hat{b}\hat{b}
    ,
\end{equation}
with $\omega_q = \sqrt{8 E_J E_C} - E_C$ and $\alpha=-E_C$.
As long as the states are in the potential well, the dominant coupling is still this ladder coupling between the adjacent levels, as will be shown later. 
However, the eigenenergies and coupling strengths deviate from the Duffing model at higher levels due to the higher-order expansion of the cosine term in~\cref{eq:transmon hamiltonian exact} \cite{Khani2009Optimal}.

An accurate effective model requires exact diagonalization up to a truncation level $N_{\rm{max}}$, which gives
\begin{equation}
    H_0 = \sum_{k=0}^{N_{\rm{max}}} \omega_k \ket{k}\bra{k},
\end{equation}
and for the charge operator:
\begin{eqnarray}\nonumber
    \hat{n}&=&\sum_{k=0}^{N_{{\rm{max}}}-1}\bigg[n_{k,k+1}\ketbra{k}{k+1}\\
            &+&\sum_{j=1} n_{k,k+2j+1}\ketbra{k}{k+2j+1}\bigg]+\rm{h.c}.
\end{eqnarray}
where $n$ denotes the corresponding matrix element.
The maximal $j$ is chosen such that $k+2j+1$ falls within the truncated levels.
Unlike the Duffing oscillator model, the $\hat{n}$ operator in the effective frame exhibits additional transitions, these non-zeros matrix elements are related to the underlying parity symmetry from the Mathieu functions, the formal solution of the Hamiltonian in~\cref{eq:transmon hamiltonian exact}.
Here, we distinguish between the ladder coupling between $\ket{k}\leftrightarrow \ket{k+1}$ and high-order couplings.
The latter are typically orders of magnitude smaller and are suppressed by the rotating wave approximation, as we will demonstrate later.

From this point, it is more convenient to express the Hamiltonian in the rotating frame defined by the transformation $R = \exp\left(-i\omega_d t\sum_{k}k\ketbra{k}{k}\right)$, where $\omega_d$ is the selected driving transition frequency. This leads to the following total Hamiltonian:
\begin{align}
\label{HT}
    \hat{H}=&\sum_{k=0}^{N_{{\rm{max}}}}\tilde{\Delta}_{k}\hat{\Pi}_{k}+\Omega(t)\cos(\omega_d t)
    \Biggr[\nonumber\\
    &\sum_{k=0}^{N_{\rm{max}}-1}\sum_{j=1}n_{k,k+2j+1}\ketbra{k}{k+2j+1}e^{-(2j+1)i\omega_d t}
    \nonumber\\
    &+
    \sum_{k=0}^{N_{{\rm{max}}}-1}n_{k,k+1}\ketbra{k}{k+1}e^{-i\omega_d t}
    +\rm{h.c}
    \Biggr]
    ,
\end{align}
where $\hat{\Pi}_{k}=\ket{k}\bra{k}$ and $\tilde{\Delta}_{k}=\omega_{k}-k\omega_{d}$ is the detuning between the $k$-th level with the $k$-th driving frequency harmonic.
In the rotating frame, all coupling terms, except for $\ket{k}\leftrightarrow\ket{k+1}$, and all the counter-rotating components oscillate rapidly.
Additionally, couplings between non-nearest-neighbour levels in a transmon ladder qudit are typically several orders of magnitude weaker, making the associated Stark shifts negligible.
Therefore, we can neglect those small contributions, leading to
\begin{equation}
\label{quditH}\hat{H}_{\rm{rwa}}=\sum_{k=0}^{N_{\rm{max}}}\tilde{\Delta}_{k}\hat{\Pi}_{k}+\frac{\Omega(t)}{2}\sum_{k=0}^{N_{{\rm{max}}}-1}(n_{k,k+1}\ketbra{k}{k+1}+\rm{h.c}).
\end{equation}
In this Hamiltonian, we recover the desired ladder coupling, albeit with renormalized eigenenergies and coupling strengths.
To target a specific ladder transition $(k, k+1)$, the drive frequency is chosen such that $\tilde{\Delta}_k=\tilde{\Delta}_{k+1}$ for the desired $k$.
The nonlinearity, captured by the remaining $\tilde{\Delta}_k-\tilde{\Delta}_j$ for $j\notin \{k, k+1\}$, permits selective driving of any transition between neighboring levels.

In addition to the Hamiltonian, we also consider the possible decoherence of the higher levels.
To use the quantum states as a qudit, we demand that they are robust against charge fluctuations, which increase exponentially up the ladder. This condition sets an upper bound on the maximal number of usable states $N_{\rm{states}}$, which a priori depends on the ratio $E_J/E_C$, and is graphed in \cref{fig:Numberstates}d as a function of anharmonicity $\alpha\approx -E_C$.
A higher $E_J$ corresponds to a deeper potential, allowing for more confined states, while a lower $E_C$ reduces charge fluctuations. However, in such a regime we also have decreases in the frequency difference between transitions, making selective driving more challenging. Instead of using the $E_J/E_C$ ratio, it is more natural to select the usable states in terms of their coherence times $T_1$ and $T_{\phi}$. In our case, as we consider fixed-frequency transmons, the main source of error will correspond to capacitive losses and dephasing due to charge fluctuations. As an optimistic forward-looking estimation, we set the upper bound of $T_{\phi}$ for the highest level $N_{\rm{states}}$ to be around $100~\mu$s such that we are able to potentially achieve gate error below $10^{-4}$ for a 10 ns gate. For the parameters we choose, this corresponds to a charge dispersion of about $10^{-3}$ GHz. The details on the calculation of the charge fluctuation and coherence times are presented in \cref{sec:decoherence}.
In principle, for certain quantum operations, decoherence could be partially mitigated by applying dynamical decoupling methods~\cite{Tripathi2025Qudit}. In this work, however, we consider more generally using quantum control shaping to speed up the operation time and reduce the irreversible effect of decoherence.

\begin{figure}
    \centering
    \includegraphics[width=\linewidth]{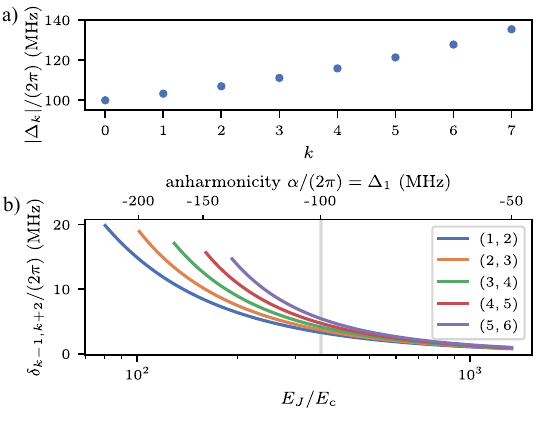}
    \caption{
    Properties of transmon qudits.
    a) Detuning between the target subspace and the leakage levels in the rotating frame. The qubit frequency and anharmonicity of the ground state are $5$~GHz and $-100$~MHz, corresponding to $E_J/E_C\approx 355$. For $k=0$, $\Delta_k=\alpha.$
    b) The small energy gap between the two leakage levels $\ket{k-1}$ and $\ket{k+2}$ for the first five transitions. This is much smaller than $\Delta_k$, leading to the energy structure shown in \cref{fig:qudit energy levels}. The grey vertical line marks the parameters used in (a).
    }
    \label{fig:ga03 sweep}
\end{figure}

\subsection{Four-level effective model}

Although the full qudit contains many levels, transitions that are detuned by more than $2\Delta_k$ from the $\ket{k}\leftrightarrow\ket{k+1}$ transition are far off-resonant in the rotating frame relative to the coupling strength and therefore have a negligible influence on the system’s dynamics.
Therefore, we focus on nearest-neighbour transitions and further simplify the model to a four-level system. This choice is validated by the numerical simulations that follow.
We define $\omega_d=\omega_{k+1}-\omega_{k} - \delta_d$, with $\delta_d$ denoting a designed small detuning between the drive frequency and the energy separation.
The special case for qubits, $k=0$, has been studied over the last decade~\cite{Motzoi2012Controlling}.
The primary control error arises from the coupling to the nearest neighboring levels, $\ket{k-1}$ and $\ket{k+2}$, as illustrated in \cref{fig:Numberstates}c.
To simplify the analysis, we truncate the Hamiltonian to a four-level subsystem, described by
\begin{equation}
\hat{H}_k^{(4)}
=
\begin{pmatrix}
\Delta_k & \frac{\lambda_{k-1}\bar{\Omega}}{2} & 0 & 0 \\
\frac{\lambda_{k-1}\Omega}{2} & \delta_d  & \frac{\lambda_{k} \bar{\Omega}}{2}  & 0 \\
0 & \frac{\lambda_{k} \Omega}{2} & 2 \delta_d  & \frac{\lambda_{k+1} \bar{\Omega}}{2}  \\
0 & 0 & \frac{\lambda_{k+1} \Omega}{2} &  3 \delta_d+\Delta_k+\delta_{k-1, k+2} \\
\end{pmatrix}
,
\label{eq:lab 4 level Hamiltonian}
\end{equation}
where $\bar{\Omega}$ denotes the complex conjugate of the pulse envelope.
For clarity, a constant identity operator has been subtracted.

The two middle levels represent the targeted transition, separated by the small drive detuning $\delta_d$.
The first and last levels correspond to the potential leakage levels $\ket{k-1}$ and $\ket{k+2}$.
An important observation is that, due to the weak nonlinearity, the energy separation between the resonant subspace and the leakage levels, $\Delta_k=\omega _{k-1}-2 \omega _k+\omega _{k+1}$, is always approximately equal to the anharmonicity $\alpha$ and only increases slightly as the levels rise. This is illustrated in \cref{fig:ga03 sweep}a, with the base case $\Delta_0=\alpha$.
The difference between the two leakage levels is given by $\delta_{k-1, k+2}=-\omega _{k-1}+3 \omega _k-3 \omega _{k+1}+\omega _{k+2}$ (see~\cref{fig:qudit energy levels}c).
For a harmonic or Duffing oscillator, it is straightforward to verify that $\delta_{k-1, k+2}$ is zero, (see Appendix \ref{sec: Leakage manifold}).
However, for a transmon oscillator, $\delta_{k-1, k+2}$ takes a small but nonzero value, as illustrated in \cref{fig:ga03 sweep}b, which is plotted as a function of $E_J/E_c$ and the anharmonicity.
The curve is truncated when the eigenstate's dispersion noise reaches $10^{-3}$ GHz, where the qudit coherence time drops below a minimum threshold (see \cref{sec:decoherence}).
Within this range, $\delta_{k-1, k+2}$ is much smaller than $\Delta_k$.

The off-diagonal coupling term in \cref{eq:lab 4 level Hamiltonian} shows a similar structure as the Duffing model. The term $\lambda_k$ denotes the renormalized drive strength between level $k$ and $k+1$, given by $\lambda_k=n_{k,k+1}/|n_{0,1}|$, which equals $\sqrt{k}$ in the Duffing approximation.
This model defined in~\cref{eq:lab 4 level Hamiltonian} holds as long as the state remains within the potential well and the eigenenergy's dispersion to charge noise is sufficiently small.

\section{Recursive DRAG pulse for qudit gates}
\label{sec:recursive DRAG qudit}

To suppress multiple leakage pathways in qudit systems, we employ the recursive DRAG method.
The challenge of addressing multiple error channels was first mentioned in Ref.~\cite{Gambetta2011Analytic}, which optimized only single-derivative DRAG parameters to balance competing leakage pathways.
The subsequent work, Ref.~\cite {Motzoi2013Improving}, introduced additional control degrees of freedom, but its main focus was on frequency selectivity in uncoupled qubit systems with linear coupling.
A similar approach with linear multi-derivative correction is also derived in \cite{Hyyppa2024Reducing}, motivated by frequency engineering.
In contrast, a more systematic extension to systems with connected transitions, such as multi-level architectures, was proposed in Ref.~\cite{Li2024Experimental}, leading to the recursive DRAG method. This approach allows for simultaneous suppression of single- and multi-photon transitions through a hierarchy of higher-order derivative corrections in a compact recursive form, but generally ignores phase and amplitude errors stemming from non-commutativity of the transitions.
The key advantage of the recursive framework is its ability to produce compact analytical pulse shapes via a structured expansion, significantly simplifying both theoretical design and experimental calibration.
Each error channel can be independently addressed, enabling modular tuning of pulse parameters.

However, the original formulation in Ref.~\cite{Li2024Experimental} was tailored to suppress fully off-resonant population transfer in a three-level system. This effectively creates quasi-adiabatic dynamics, but does not take into account typical diabatic effects present in most gates, such as the qudit system presented here. In fact, there are two quasi-diabatic transitions that are present in the qudit error model, namely the resonant $\ket{k} \leftrightarrow\ket{k+1}$ and the near-degenerate $\ket{k-1} \leftrightarrow\ket{k+2}$, especially important for high power (see \cref{fig:Numberstates}c). The interplay between the off-resonant and on-resonant dynamics thereby entails several new error channels coming from large non-commutative effects, remaining leading order in the expansion, and necessitating accounting for the phase error, which is ignored in Ref.~\cite{Li2024Experimental} because it commutes with the adiabatic dynamics.
Moreover, the presence of these distinct leakage levels in our model offers an opportunity to systematically study calibration strategies and assess how each DRAG coefficient influences specific leakage pathways.
In the following, we first analyze the error budget associated with such qudit gates and present an enhanced recursive DRAG protocol designed for this control challenge.
Our method concurrently suppresses up to four off-resonant transitions, while maintaining high-fidelity coupling between the target states $\ket{k}$ and $\ket{k+1}$.

\begin{figure}
    \centering
    \includegraphics[width=\linewidth]{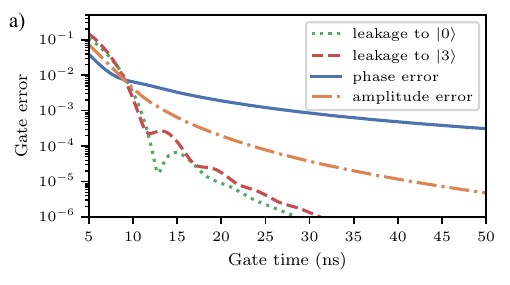}
    \includegraphics[width=\linewidth]{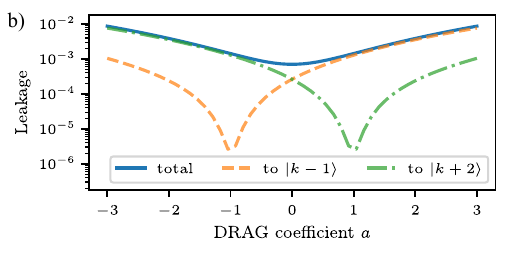}
    \caption{Control error in driving ladder transitions in a transmon qudit.
    a) Estimated error budget of driving a $\ket{1}\leftrightarrow\ket{2}$ $\pi$ rotation using a Hann pulse with an anharmonicity of $\alpha/(2\pi)=-200$~MHz. The definitions of different errors are given in Appendix \ref{sec:error defintion}.
    b) The leakage error calculated via Eq.~\ref{eq:leakage} in a regime where the single-derivative DRAG faces limitations and offers no improvement, even with an optimized DRAG coefficient.
    Parameters used are $\lambda_{k-1}=\lambda_k=\lambda_{k+1}=1$, $\delta_{k-1,k+2}=0$ and $\tfinal=15$ns. The pulse is a single-derivative DRAG shape $\Omega-ia\dot{\Omega}/\Delta_k$, with $\Omega$ the standard Hann pulse.}
    \label{fig:error budget}
\end{figure}

\subsection{Error budget}
To identify the dominant sources of error in qudit gate operations, we begin by analyzing the error budget associated with driving transitions in a transmon qudit.
For the lowest two levels in a transmon, $\ket{0}$ and $\ket{1}$, the system reduces to the well-studied transmon qubit (\cref{fig:qudit energy levels}b), well-established techniques like DRAG have been developed to suppress leakage, particularly to $\ket{2}$.
In higher-level transitions, however, leakage and control errors become more complex due to residual couplings to additional non-computational states (\cref{fig:qudit energy levels}c).
For a general $\ket{k} \leftrightarrow \ket{k+1}$ gate, adjacent states $\ket{k-1}$ and $\ket{k+2}$ introduce errors such as leakage and Stark shifts, which grow more significant at shorter gate durations.

To quantify these contributions, we consider the Hann envelope,
\begin{equation}
    \Omega_{\rm{Hann}}(t) = \sin \left[\frac{\pi t}{\tfinal}\right] ^ 2,
    \label{eq:Hann}
\end{equation}
a commonly used smooth pulse of duration $t_{\rm{f}}$ with a narrow spectral bandwidth that usually performs reasonably well.
The corresponding error contributions are shown in \cref{fig:error budget}a, with definitions provided in Appendix~\ref{sec:error defintion}.
As evident from the figure, both leakage to $\ket{k-1}$ and $\ket{k+2}$ must be addressed.
Additionally, residual couplings introduce Stark shifts that alter the resonance condition and effective coupling strength.
To ensure high-fidelity rotations within the ${\ket{k}, \ket{k+1}}$ subspace, it is essential to suppress not only leakage but also phase and amplitude errors.
This is different from previous studies, where phase error commutes with the desired dynamics and is automatically removed under the standard calibration routine~\cite{Li2024Experimental}.
As shown in \cref{fig:error budget}a, the latter dominates at long gate durations, while leakage errors increase sharply as the gate is sped up.
Moreover, using more qudit levels typically requires reducing the qudit nonlinearity $\Delta_k$ to suppress charge noise (\cref{fig:qudit energy levels}d), further complicating fast gate design.

In the standard qubit scenario, such errors are suppressed using the DRAG pulse~\cite{Motzoi2009Simple}:
\begin{equation}
\Omega = \Omega_{\rm{Hann}} - \im a \frac{\dot{\Omega}_{\rm{Hann}}}{\Delta},\label{eq:single-drag}
\end{equation}
where $\Delta$ is the detuning to the leakage level, and $a$ is a tunable coefficient to compensate for imperfect Hamiltonian modelling and higher-level corrections.
An additional detuning $\delta_d$ is often included to correct phase accumulation between driven levels~\cite{Chen2016Measuring}.

For transitions involving higher states in qudit(\cref{fig:qudit energy levels}c), however, this single-derivative correction becomes insufficient. 
The standard DRAG form in \cref{eq:single-drag} cannot simultaneously suppress both, as the associated detunings are approximately equal in magnitude but opposite in sign: $E_{\ket{k}} - E_{\ket{k-1}} \approx -(E_{\ket{k+2}} - E_{\ket{k+1}})$.
As a result, the spectral notch introduced by the single derivative cannot align with both frequencies, leading to negligible improvement in gate performance, as seen in \cref{fig:error budget}b. This limitation is intrinsic to transitions with $k \ge 1$, due to the level spacing in transmon ladders.

\subsection{General DRAG correction for a $n$-photon transition}
To address this limitation, an effective strategy involves incorporating higher-order derivative terms~\cite{Motzoi2013Improving,Li2024Experimental}. This approach can be interpreted as a superadiabatic transformation~\cite{Deschamps2008Superadiabaticity}, wherein a second adiabatic frame is derived, enabling the introduction of new time-dependent control functions proportional to the second derivative of the original drive shape.
In the presence of multiple leakage levels, DRAG corrections can be tailored for each leakage coupling and chained together. In the following, we first present the general formulation and then derive the specific solution to this problem.

For an $n$th-order coupling $\Omega^n/\Delta_{\mathrm{eff}}^{n-1}$ between two levels separated by $\Delta$, the Hamiltonian in the two-level subspace is given as 
\begin{equation}
\label{eq:two-level Hamiltonian}
\hat{H}=
-\frac{\Delta}{2}\hat{\sigma}^z_{jk}
+\left(
\frac{\Omega^n}{\Delta_{\mathrm{eff}}^{n-1}} \frac{\hat{\sigma}^+_{jk}}{2}+\rm{h.c.}
\right)
.
\end{equation}
where \(\hat{\sigma}^z_{jk}={\left(\ket{k}\bra{k}-\ket{j}\bra{j}\right)}\) and \(\hat{\sigma}_{jk}^+=\ket{k}\bra{j}\).
The symbol $\Delta_{\mathrm{eff}}$ represents an effective prefactor that renormalizes the coupling strength.
Assuming $\Omega^n/\Delta_{\mathrm{eff}}^{n-1} \ll \Delta$, we perform a perturbative expansion with the antihermitian generator $\hat{S}(\tilde \Omega)=\frac{\tilde{\Omega}^n}{2\Delta\Delta_{\mathrm{eff}}^{n-1}}\hat{\sigma}^+_{jk}-\rm{h.c}$.
The time-dependent frame transformation is given as 
\begin{equation}
\hat{H}'(\Omega)= \hat{V}(\tilde{\Omega})\hat{H}(\Omega)\hat{V}^{\dagger}(\tilde{\Omega})+\im \dot{\hat{V}}(\tilde{\Omega})\hat{V}^{\dagger}(\tilde{\Omega})
\label{eq:qudit frame transformation}
\end{equation}
with $ \hat{V}(\tilde{\Omega})=e^{\hat S(\tilde{\Omega})}$. This transformation yields
\begin{align}
    \hat{H}'(\Omega) = &i\dot{\hat{S}}(\tilde \Omega) + \hat{H}(\Omega) + [\hat{S}(\tilde \Omega), \hat{H}(\Omega)] + \cdots\\ \nonumber
    \approx &
    -\frac{\Delta}{2}\hat{\sigma}^z_{jk}
    +
    \frac{1}{\Delta_{\mathrm{eff}}^{n-1}}
    \left(
    \Omega^n
    -
    \tilde{\Omega}^n
    +
    i\frac{\dd}{\dd t}
    \frac{\tilde{\Omega}^n}{\Delta}
    \right) \frac{\hat{\sigma}^+_{jk}}{2}
    \nonumber\\
    &+\rm{h.c.}
    ,
    \nonumber
\end{align}
where we keep only the leading-order perturbation.
Following from the equation above, the DRAG pulse is given by
\begin{equation}
    \Omega^n
    =
    \tilde{\Omega}^n
    -
    i\frac{\dd}{\dd t}
    \frac{\tilde{\Omega}^n}{\Delta}
    .
    \label{eg:general DRAG}
\end{equation}
Therefore, we can derive a drive pulse $\Omega$ resistant to this error based on the initial shape $\tilde{\Omega}$ and its derivative.

To ensure that the unitary evolution remains consistent under the frame transformation in \cref{eq:qudit frame transformation}, it is crucial that the generator $\hat{S}$ vanishes at the beginning and end of the evolution.
To achieve this, we use the following initial pulse shape:
\begin{equation}
\Omega_{\text{I}}(t) = \Omega_{\textnormal{max}}\left[\frac{1}{16}\cos \left[6\pi\frac{t}{\tfinal}\right]-\frac{9}{16}\cos \left[2\pi
 \frac{t}{\tfinal}\right]+\frac{1}{2}\right],
\label{eq:qudit DRAG initial pulse}
\end{equation}
with $\Omega_{\textnormal{max}}$ the drive amplitude.
This pulse is designed to ensure that the function and its derivatives up to fourth order vanish at the start and end of the gate, as required for our frame transformation.
However, without any correction term, it has more spectral spreading than the Hann pulse and exhibits worse performance.

\subsection{First-order (linearized) solution for qudits}

To manage the two different leakage channels with opposite energy gaps as shown in \cref{fig:qudit energy levels}c and \cref{fig:error budget}b, two degrees of freedom are required.
The leading-order leakage errors in \cref{eq:lab 4 level Hamiltonian} is associated with the ladder couplings between $\ket{k-1}\leftrightarrow\ket{k}$ and $\ket{k+1}\leftrightarrow\ket{k+2}$.
Both of these are first-order transitions with $n=1$ in \cref{eq:two-level Hamiltonian}.
To address the two errors, two DRAG corrections can be introduced recursively~\cite{Motzoi2013Improving,Li2024Experimental} as
\begin{align}
    \Omega = \Omega_1 -\im\frac{\dot{\Omega}_1}{\Delta_{L}},
    \label{eq: DRAG2 corrections1}
    \\
    \Omega_1 = \Omega_2 -\im\frac{\dot{\Omega}_2}{\Delta_{H}},
    \label{eq: DRAG2 corrections2}
\end{align}
where $\Delta_{H}=E_{\ket{k+2}}-E_{\ket{k+1}}$ and $\Delta_{L}=E_{\ket{k}}-E_{\ket{k-1}}$ are energy gaps with respect to the upper and lower adjacent levels.
For $\Omega_2$ we use $\Omega_{\text{I}}$ in~\cref{eq:qudit DRAG initial pulse}.
The detailed derivation based on perturbation theory is provided in \cref{sec:derivation DRAG}.
Each of these expressions is designed to address one leakage pathway, and their order is interchangeable due to the linearity of derivatives.

This is different from the high-order perturbative solution proposed in Ref.~\cite{Gambetta2011Analytic}, where no second derivatives were introduced and the result is only a compromise between different errors.
This recursive formulation suppresses both errors simultaneously to the leading order and can be extended with additional correction terms if more ancillary levels are involved ~\cite{Motzoi2013Improving}.
Semiclassically, it can be understood as engineering two zero points on the classical spectrum of the pulse.
We refer to this DRAG pulse as the DRAG2 pulse.
Notably, for a weakly nonlinear oscillator where $\Delta_{L}\approx-\Delta_{H}$, the imaginary part of the correction becomes small, and the real part dominates:
\begin{equation}
    \Omega \approx \Omega_{\text{I}} + \frac{\ddot{\Omega}_{\text{I}}}{\Delta_{L}^2} \approx \Omega_{\text{I}} + \frac{\ddot{\Omega}_{\text{I}}}{\Delta_{H}^2}
    .
\end{equation}

Apart from the leakage error, two other errors, the phase and amplitude errors, must also be addressed to achieve the desired rotation.
For a typical qubit operation between $\ket{0} \leftrightarrow \ket{1}$, the phase error comes from both the Stark shift caused by the $\ket{2}$ state and the non-commutativity of the imaginary DRAG correction term.
For transitions involving higher levels, the Stark shift is influenced by both the higher and lower adjacent levels. Because the phase accumulation on the states $\ket{k}$ and $\ket{k+1}$ have the same sign, the overall accumulated phase error in this two-level transition is smaller compared to driving $\ket{0} \leftrightarrow \ket{1}$~\cite{Lucero2010Reduced}.
Experimentally, this small phase error is often mitigated by applying a constant detuning to the drive ~\cite{Chen2016Measuring}.
The correction of the drive shape also affects the rotation angle.
To compensate for this, a small correction term needs to be added, $\Omega_2 \leftarrow \Omega_2 + \Omega_{\text{amp}}$.
Similar to the phase correction, this amplitude error can also be approximately mitigated by calibrating the maximal drive amplitude $\Omega_{\max}$.

\subsection{Second-order solution for qudits}
Although the two couplings between $\ket{k-1}\leftrightarrow\ket{k}$ and $\ket{k+1}\leftrightarrow\ket{k+2}$ are suppressed by the DRAG2 correction, under a strong drive the higher-order transitions between $\ket{k-1}\leftrightarrow\ket{k+1}$ and $\ket{k}\leftrightarrow\ket{k+2}$ may also play a role.
These second-order transitions arise from diagonalizing the direct ladder couplings and are proportional to $\Omega^2$ (see \cref{sec:derivation DRAG}).
Following the general DRAG expression in \cref{eg:general DRAG}, this leads to the chained second-order correction:
\begin{align}
    \Omega_2 = \sqrt{\Omega_3^2-\im \frac{2\Omega_3\dot{\Omega}_3}{\Delta_{H}}}
    \label{eq: DRAG4 corrections1}\\
    \Omega_3 = \sqrt{\Omega_4^2-\im \frac{2\Omega_4\dot{\Omega}_4}{\Delta_{L}}}
    \label{eq: DRAG4 corrections2}
\end{align}
where $\Omega_4$ is again taken from $\Omega_{\text{I}}$ in \cref{eq:qudit DRAG initial pulse}.
We refer to this pulse, combined with the two corrections in \cref{eq: DRAG2 corrections1,eq: DRAG2 corrections2} as the DRAG4 pulse.

The second-order corrections introduced above commute with each other but do not commute with the first-derivative corrections.
It is important to note that in the recursive DRAG formulation, the second-order correction is applied first to the initial pulse. This ensures that the dynamics in the final effective frame are governed by the initial pulse. This ordering is the reverse of the order of perturbation.

\subsection{Performance benchmarking}

To demonstrate the performance of the recursive DRAG pulse, we simulate the time evolution for various gate durations and compare different drive schemes.
We use the analytically derived DRAG pulse while numerically calibrating constant detuning $\delta_d$ and amplitude $\Omega_{\text{max}}$.
The simulation is performed using the full Hamiltonian, truncated at $N_{\max}$, which is much larger than the qudit size.
The average gate fidelity is calculated as~\cite{Pedersen2007Fidelity}
\begin{equation}
    \label{eq:two level fidelity}
    F [\hat{\mathcal{U}}_Q] =
    \frac{\Tr [\hat{\mathcal{U}}_Q \hat{\mathcal{U}}^{\dagger}_Q]}{d(d+1)}
    + \frac{ \left|\Tr [\hat{\mathcal{U}}_Q \hat{\mathcal{U}}_I^{\dagger}] \right|^2}{d(d+1)},
\end{equation}
with $\hat{\mathcal{U}}_Q$ representing the truncated unitary within the two-level subspace for the targeted transition and $\hat{\mathcal{U}}_I$ the ideal $\pi$ and $\pi/2$ rotation gates.
Note that this fidelity only includes deviations in the gate quality within the two-level subspace and error leakages from the target states $\ket{k}$ and $\ket{k+1}$.
Error dynamics that may occur on other levels are discussed in \cref{sec:other error}.

\begin{figure}
    \centering
    \includegraphics[width=\linewidth]{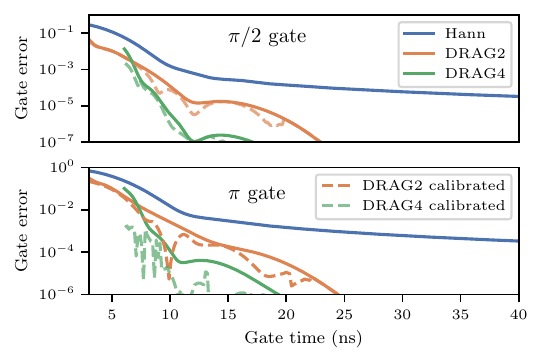}
    \caption{Gate infidelity as a function of duration for different drive schemes driving the $\ket{1} \leftrightarrow \ket{2}$ transition, with $\alpha/(2\pi)=-200$~MHz and $\omega_{10}/(2\pi)=5$~GHz. The DRAG2 pulse is defined in \cref{eq: DRAG2 corrections1,eq: DRAG2 corrections2} and the DRAG4 pulse in \cref{eq: DRAG4 corrections1,eq: DRAG4 corrections2}.
    The solid line represents the analytically derived DRAG pulse and the dashed line represents gate error with calibrated DRAG coefficients.
    }
    \label{fig:gate fidelity vs time}
\end{figure}

In \cref{fig:gate fidelity vs time}, we compare the gate fidelity between standard Hann pulse, DRAG2 and DRAG4 pulses, for $\pi$ and $\pi/2$ gates on $\ket{1} \leftrightarrow \ket{2}$.
For the Hann pulse defined in Equation 9, we set the integral of the pulse such that it equals the required gate rotation angle.
For short gate durations, below 25~ns, each successive correction improves fidelity by one to two orders of magnitude.
For longer gate times, the error is primarily dominated by phase and amplitude errors, which are suppressed by the constant detuning and amplitude optimization for the DRAG pulses.
For gate durations shorter than 7~ns, DRAG4 data is omitted, as the pulse amplitude becomes several times larger than the base envelope, violating the perturbative assumptions and is impractical for experimental implementation.
This improvement can also be examined by fixing a target fidelity and examining the minimum gate time required to achieve it.
For instance, with a target fidelity of $10^{-4}$, the DRAG pulse reduces the gate duration to 10~ns from 30~ns for a $\pi/2$ gate, and from more than 40~ns to 15~ns for a $\pi$ gate.

\begin{figure}
    \centering
    \includegraphics[width=\linewidth]{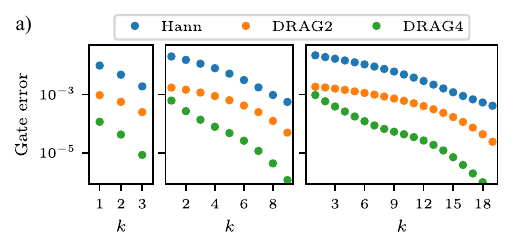}
    \includegraphics[width=\linewidth]{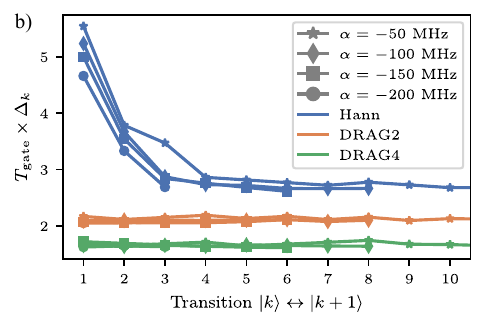}
    \caption{Controlling the ladder transitions in a transmon qudit. a) The $\pi/2$ gate error for different ladder transitions $\ket{k}\leftrightarrow\ket{k+1}$ for $\alpha/(2\pi)=-200, -100, -50$~MHz using a fixed gate duration of 8, 15 and 30~ns respectively. These values of anharmonicity correspond to $E_J/E_C\approx$100, 355 and 1331, respectively. b) The minimum gate time required to achieve infidelity of $10^{-4}$ for different drive schemes and different hardware parameters for a $\pi/2$ gate. The gate duration is multiplied by the corresponding leakage level separation $\Delta_k$, resulting in overlapping outcomes across different hardware parameters and qudit levels.}
    \label{fig:ladder fidelity vs n}
    \label{fig:minimum gate time}
\end{figure}

Generalizing the analysis to arbitrary $\ket{k}\leftrightarrow\ket{k+1}$ transitions, we apply the same DRAG construction and repeat the benchmarking for different $k$ values.
\Cref{fig:ladder fidelity vs n}a shows the fidelity improvement for three different anharmonicities $\alpha$ and gate times.
As the anharmonicity decreases, the system more closely resembles a linear oscillator, allowing more levels to be used as quantum registers without significant coupling to environmental noise.
However, the energy difference between each level, roughly proportional to the anharmonicity, also decreases.
Therefore, we increase the gate time proportionally, inversely to the reduced anharmonicity. 
The results indicate that the improvements provided by the DRAG corrections for general $\ket{k}$ transitions with varying anharmonicity are analogous to those observed for $\ket{1}\leftrightarrow\ket{2}$ in \cref{fig:gate fidelity vs time}. This also verifies that the four-level effective model is well suited for studying the transmon ladder transition across different levels. In addition, we observe that control errors decrease as the level $k$ increases, mainly because the leakage coupling to upper and lower levels becomes more symmetric as the level goes up.
As the two leakage couplings $\ket{k-1}\leftrightarrow\ket{k}$ and $\ket{k+1}\leftrightarrow\ket{k+2}$ become more symmetric, the phase error is reduced for higher levels, as also observed in Ref.~\cite{Lucero2010Reduced}.

To further characterize the control of different ladder transitions, we compute the minimal gate duration achievable for an infidelity threshold of $10^{-4}$, as depicted in \cref{fig:minimum gate time}b.
To capture the universality of the pulse solutions, we normalize the time by the energy separation $\Delta_k$ for each ladder transition. Remarkably, we see that when comparing different transition indices $k$, and comparing different values of the anharmonicity, all the values collapse horizontally on the same line for the DRAG family of pulses. When we use DRAG4 pulses instead of DRAG2, we remove two additional weak, 2-photon transitions, and these collapse to a yet shorter minimum time line (related to a quantum speed limit for the particular choice of pulse), with apparently even stronger overlap for different parameters. 
However, this is not the case for the standard Hann pulse, which exhibits a strong dependence on both the anharmonicity (or equivalently \(E_J/E_C\)) and the chosen level index. This is due to phase and amplitude errors introduced by asymmetric coupling to neighboring levels. Eliminating those errors, together with the suppression of leakage of different orders, reduces the system to a universal qubit-like model, whose behavior is independent of the specific nature of the suppressed leakage channels.

\subsection{Calibration of the DRAG coefficients}

\begin{figure*}
    \centering
    \includegraphics[width=\linewidth]{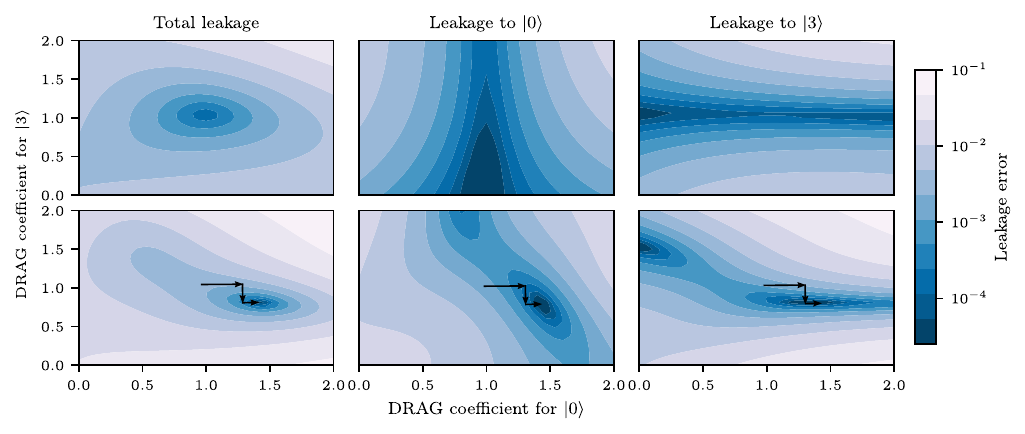}
    \caption{Calibration of the DRAG2 pulse. Plotted is the leakage error as a function of the DRAG coefficients for DRAG2 pulses $a_{1,0}$ and $a_{1,3}$ for a $\ket{1}\leftrightarrow \ket{2}$ $\pi$ gate with $\alpha/2\pi = -200$~MHz, with a gate duration of 15ns (upper row) and 10ns (lower row). Amplitude and detuning are optimized for each pair of coefficients. While the analytical prediction sets both coefficients to one, the lower row shows deviations due to additional coherent interference effects.
    }
    \label{fig:calibration DRAG2}
\end{figure*}

In practice, due to model inaccuracy, the DRAG pulse often needs to be calibrated.
A major advantage of the DRAG framework is its modularity and ease of experimental calibration, features that are naturally inherited by the recursive DRAG approach. In this formalism, each error mechanism is addressed by a corresponding derivative correction term, each with its own tunable coefficient $a_{n,l}$. The generalised recursive expression takes the form:
\begin{equation}
    \Omega^n
    =
    \tilde{\Omega}^n
    -
    ia_{n,l}\frac{\dd}{\dd t}
    \frac{\tilde{\Omega}^n}{\Delta_l}
    .
\end{equation}
where $n$ denotes the expansion order and $l$ the target leakage level.
For DRAG2 and DRAG4 schemes, this yields two and four independently tunable coefficients, respectively. Ideally, each coefficient targets a distinct error channel.

While the analytical prediction sets all $a_{n,l} = 1$, deviations can arise even in simulation due to coherent interference effects among leakage pathways.
As shown in~\cref{fig:gate fidelity vs time}, the DRAG2 pulse with default coefficients already performs well in most regimes, but targeted calibration can further enhance fidelity for specific gate durations.

\cref{fig:calibration DRAG2} shows the leakage error landscape for a $\ket{1} \leftrightarrow \ket{2}$ $\pi$ gate, plotted over the DRAG2 parameters $a_{1,0}$ and $a_{1,3}$ for two gate durations.
At $t_f = 15$~ns, the analytical values ($a_{1,0} = a_{1,3} = 1$) are nearly optimal, and each coefficient predominantly affects leakage to its associated level: $a_{1,0}$ governs leakage to $\ket{0}$, and $a_{1,3}$ to $\ket{3}$. 
Off-target tuning, such as adjusting $a_{1,3}$ when measuring the population of $\ket{0}$, has minimal effect and typically converges toward zero, as smaller corrections reduce unwanted spectral content.

At shorter gate times (e.g., $t_f = 10$~ns), optimal coefficients may deviate from their analytic values due to interference effects, allowing for further cancellation of leakage errors.
Nevertheless, the parameter-to-channel mapping remains approximately valid.
This observation enables a practical iterative calibration strategy: one may first optimize $a_{1,0}$ by minimizing population in $\ket{0}$, then adjust $a_{1,3}$ to suppress leakage to $\ket{3}$, and finally refine $a_{1,0}$ if needed. The progression of this calibration procedure is illustrated by the arrow markers in the plots.

The DRAG4 case involves a more complex interplay, with multiple coefficients contributing to the same leakage level. As a result, calibration may require joint optimization over two parameters. We discuss this in detail in Appendix~\ref{sec: DRAG4 calibration}.

\section{Error beyond the targeted two-level subspace}
\label{sec:other error}
In the previous analysis, we focused on the relevant two-level subspace and computed the average gate fidelity of driving a $\pi$ or $\pi/2$ rotation, using \cref{eq:two level fidelity}.
For a target transition between $\ket{k}\leftrightarrow\ket{k+1}$, this error model includes leakage from the target subspace to $\ket{k-1}$ and $\ket{k+2}$ and the corresponding phase and amplitude errors.
To use it as the building block for universal qudit computational gates, we also need to study its effect on all the $K$ qudit states.

\subsection{Phase error beyond the two target levels.}
As discussed above, the Stark shift accumulates phases on the affected subspace. The optimized detuning fixes the difference between the phase on $\ket{k}$ and $\ket{k+1}$.
However, a phase shift still exists between the subspace and other energy levels.
For the target ladder transition, this phase shift is merely a global phase, but for a $K$-level qudit, it becomes relevant and must be accounted for. 

Fortunately, this phase mismatch can be easily calibrated by applying virtual phase gates to each untargeted level~\cite{Morvan2021Qutrit}. The accumulated phase is calibrated by using a phase-amplification technique. For an operation $\text{RX}^{(k,k+1)}_{\pi/2}$, the state $(\ket{k} + \ket{j}) / \sqrt{2}$ is prepared, where $\ket{j}$ is the state not addressed by the gate. The gate is then applied $8n$ times, followed by a rotation of the system back using $\text{RY}^{(k,k+1)}_{\pi/2}$, similar to a Ramsey experiment. The accumulated phase is then measured on the state $\ket{j}$ and corrected for future use. Virtual phase gate construction is described in Appendix \ref{sec:qudit decompostion}.

\subsection{Leakage on \(\ket{k+2}\leftrightarrow\ket{k+3}\)}
The DRAG pulse we studied primarily targets leakage involving the target subspace, i.e., leakage from the two target levels to the nearest neighbours, which are separated by approximately $\Delta_k$ in the rotating frame.
Under a very strong drive, a small population transfer may also appear between nearby states such as $\ket{k+2}\leftrightarrow\ket{k+3}$, which are not directly driven.
The level separation between them is about $2|\Delta_k|$.
Due to this large separation, the unwanted transition is much smaller, however, it might become non-negligible ($>10^{-4}$) if a DRAG4 pulse is used for a short gate time.

\subsection{Three-photon leakage $\ket{k-1} \leftrightarrow \ket{k+2}$}
\begin{figure}
    \centering
    \includegraphics[width=\linewidth]{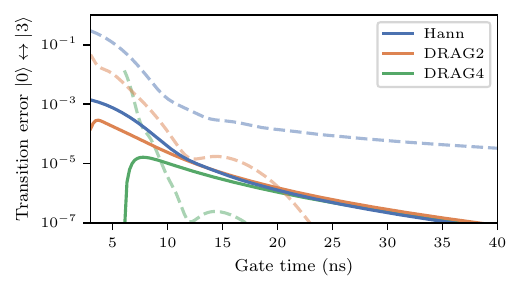}
    \caption{The three-photon transition error under different drive schemes. The solid lines represent the $\ket{0}\leftrightarrow\ket{3}$ error as a function of the gate time for different drive schemes for a $\pi/2$ gate. The dashed lines represent the gate errors computed for the target subspace for each drive scheme, same as those shown in \cref{fig:gate fidelity vs time}a. The parameters used are also the same. }
    \label{fig:leakage with 03}
\end{figure}
Another small error that we have not discussed is the three-photon transition between $
\ket{k-1}\leftrightarrow\ket{k+2}$.
As illustrated in \cref{fig:qudit energy levels}c, this third-order transition is induced by off-resonant ladder couplings and is proportional to $\Omega^3$.
Although the absolute value of this error is accordingly small, due to the very small value of $\delta_{k-1, k+2}$ in a nonlinear oscillator, it may still introduce a non-negligible error after DRAG corrections.
This error probability, defined by
\begin{equation}
    \mathcal{L}_{k-1, k+2} = \frac 14 \left(|\hat{\mathcal{U}}_{k-1, k+2}|^2 + |\hat{\mathcal{U}}_{k+2, k-1}|^2\right)
\end{equation}
is plotted in \cref{fig:leakage with 03} for $k=1$.
Targeting a gate error of $10^{-4}$, we see from the plot that this error is generally lower than the two-level gate error calculated in \cref{sec:recursive DRAG qudit}, and hence does not pose a significant obstacle.

\section{Conclusion and discussion}
\label{sec:qudit conclusion}

Using a universal pulse construction, we have shown how coherent errors can be drastically suppressed in qudit ladder systems. Moreover, they completely predict a universal behaviour whereby the minimum gate time to achieve $10^{-4}$ error can be very accurately calculated, irrespective of the details of the exact energy structure of the nearby surrounding levels.

The method adopts a recursive structure to simultaneously suppress multiple leakage errors.
Despite the simple form, the performance benchmarking highlights the substantial error reduction achieved, enabling faster gates and consequently reducing decoherence.
This method not only improves gate fidelity for the nonlinear oscillator but also offers a framework that can be adapted to other qudit systems beyond the specific model studied here.

Our results offer valuable insights into the relationship between the number of qudit levels that can be utilized for universal computation, and the corresponding gate time required to achieve a specific fidelity threshold using analytical DRAG pulses. 
For specific device parameters, the analytical pulses can be further optimized to harness additional coherent interference for better performance or to enhance robustness against system uncertainties.
Residual errors beyond those addressed in this work may also be suppressed by further including higher-order derivatives in the DRAG framework or by introducing a weak off-resonant correction drive~\cite{Chiaro2025Active}, which we leave for future investigation.

For advanced hardware with high bandwidth waveform generators, where multiplexed frequency is possible, it would be advantageous to implement non-overlapping ladder transitions in parallel, further improving the efficiency and scalability of qudit-based quantum computing.
However, drive-induced Stark shifts and inter-channel crosstalk must be carefully accounted for, requiring additional calibration to ensure accurate and high-fidelity gate operations in parallel settings.

\appendix

\section{Universality of the ladder transition}
\label[appendix]{sec:qudit decompostion}
In this section, we show that any $K$-dimensional qudit unitary $\hat{U}$ can be decomposed into $\pi/2$ gates between adjacent levels (ladder gates) $\ket{k}$, $\ket{k+1}$ and virtual phase gates. 

\subsection{Decomposition of arbitrary unitary to Givens rotation}
In the following, we show that arbitrary qudit unitaries $SU(d)$ can be decomposed into a sequence of Givens rotations and a diagonal phase matrix.
We follow the QR decomposition similar to Ref.~\cite{Ramakrishna2000Explicit,Brennen2005Criteria} and decompose the unitary by progressively eliminating the left bottom part of the unitary matrix.
An upper triangular matrix which is unitary is easily proved to be a diagonal matrix.

A unitary Givens rotation between two levels $j, l$ is defined as
\begin{equation}
\hat{G}(\gamma, \varphi) =
\left(
\begin{array}{ccccc}
\ddots\\
&\cos{\frac{\gamma}{2}} &  \dots  & -\im e^{\im \varphi} \sin{\frac{\gamma}{2}}\\
&\vdots & \ddots & \vdots\\
&-\im e^{-\im \varphi} \sin{\frac{\gamma}{2}} & \dots  & \cos{\frac{\gamma}{2}}\\
&&&&\ddots
\end{array}
\right)
.
\end{equation}
The entries not explicitly defined are filled with identities, i.e., one in the diagonal entries and zero otherwise.

The QR decomposition eliminates each column from left to right and for each row from bottom to the diagonal.
This ensures that the eliminated entries will remain zero in later steps.
There are in total $M=(K-1)K/2$ Givens rotations.
For the $m$-th Givens rotation $G^{(m)}$ designed to eliminate the entry $j,l$, the parameters are recursively defined by
\begin{align}
\tan{\gamma_{m}} &= 2{|U^{(m)}_{j,l}/U^{(m)}_{j-1,l}|},\\
\varphi_{m} &= \pi/2 + \textnormal{arg}(U^{(m)}_{j-1,l}) - \textnormal{arg}(U^{(m)}_{j,l}),
\end{align}
where $\hat{U}^{(m)}$ is the remaining unitary after eliminating the first $m$ entries, i.e., $\hat{U}^{(m)}=\hat{G}^{(m-1)}\hat{G}^{(m-2)}\cdots\hat{G}^{(1)}\hat{U}$.

Note that the Givens rotation applied above is always a rotation between two adjacent levels $j, j-1$, which can be directly implemented by ladder coupling.
To further simplify the native gate set, it is common to use the ZXZXZ decomposition~\cite{McKay2017Efficient} designed for a qubit, which decomposes an arbitrary Givens rotation into three RZ gates and two $\pi/2$ gates
\begin{align}
&\hat{G}(\gamma, \varphi) =\text{RZ}({-\varphi -\frac{\pi }{2}})\text{RX}_{\frac{\pi }{2}}\text{RZ}({\pi -\gamma })\text{RX}_{\frac{\pi }{2}}\text{RZ}({\varphi -\frac{\pi }{2}})
,
\end{align}
where the subscript denotes the rotation angle. The diagonal matrix after the QR decomposition can also be easily written as a series of RZ gates. However note that the DRAG2 and DRAG4 pulses derived in the main text can also be used with any rotation and phase angle.

\subsection{Virtual phase gate in a qudit}
The above decomposition resolves arbitrary unitary to $\pi/2$ gates and RZ gates.
The former is the main focus of the main text.
Here, we show that the RZ gate between two arbitrary qudit levels, defined by
\begin{align}
\textnormal{RZ}(\lambda)
=
\left(
\begin{array}{cc}
 e^{-i\lambda/2} & 0 \\
 0 & e^{i\lambda/2}\\
\end{array}
\right),
\end{align}
can be implemented by adjusting a constant phase of the drive~\cite{McKay2017Efficient}.
This virtual phase implementation will significantly reduce the duration of the circuit.
In contrast to the case of qubits, since there are $K-1$ ladder transitions, we need to track the accumulated phase for each drive, which we denote as $\theta_k$.

A rotation between $\ket{k}$ and $\ket{k+1}$ is implemented by a Hamiltonian
\begin{equation}
\hat{H}_{\Omega}^{(k)}(\theta)
=
\left(
\begin{array}{cc}
 0 & \frac{1}{2} e^{\im \theta_k } \Omega(t) \\
 \frac{1}{2} e^{-\im \theta_k } \Omega(t) & 0\\
\end{array}
\right),
\end{equation}
where $\Omega(t)$ is the time dependent drive amplitude and $\theta$ a constant phase of the drive.
This angle $\theta$ adjusts the axis in the XY plane, around which the rotation is performed.
The corresponding unitary evolution is denoted by $\hat{\mathcal{U}}^{(k)}_{\Omega}(\theta)=e^{-i\hat{H}^{(k)}_{\Omega}(\theta)t}$. It is then straightforward to show that 
\begin{equation}
\textnormal{RZ}^{(k,k+1)}(\phi) \hat{\mathcal{U}}^{(k)}_{\Omega}(\theta) = \hat{\mathcal{U}}^{(k)}_{\Omega}(\theta-\phi)\textnormal{RZ}^{(k,k+1)}(\phi).
\end{equation}
This relation indicates that by phase shifting the drive $\Omega$ by $-\phi$,  the RZ phase gate can be effectively moved to the end of the gate operation. 
Since it appears at the end, it can be neglected, as measurements only capture state populations.

The above is enough for qubit operation.
For qudit operation, however, the presence of other computational levels has to be taken into consideration.
Therefore, we need to consider an RZ gate between arbitrary two levels and obtain the following expressions
\begin{align}
&\textnormal{RZ}^{(j,k)}(\phi) \hat{\mathcal{U}}^{(k)}_{\Omega}(\theta) = \hat{\mathcal{U}}^{(k)}_{\Omega}(\theta+\phi/2)\textnormal{RZ}^{(j,k)}(\phi),\\
&\textnormal{RZ}^{(k,l)}(\phi) \hat{\mathcal{U}}^{(k)}_{\Omega}(\theta) = \hat{\mathcal{U}}^{(k)}_{\Omega}(\theta-\phi/2)\textnormal{RZ}^{(k,l)}(\phi),\\
&\textnormal{RZ}^{(j,k+1)}(\phi) \hat{\mathcal{U}}^{(k)}_{\Omega}(\theta) = \hat{\mathcal{U}}^{(k)}_{\Omega}(\theta-\phi/2)\textnormal{RZ}^{(j,k+1)}(\phi),\\
&\textnormal{RZ}^{(k+1,l)}(\phi) \hat{\mathcal{U}}^{(k)}_{\Omega}(\theta) = \hat{\mathcal{U}}^{(k)}_{\Omega}(\theta+\phi/2)\textnormal{RZ}^{(k+1,l)}(\phi),
\end{align}
with $j<k$ and $l>k+1$.
Notice that the adjusted phase is reduced by half because only one of the levels overlaps between RZ and the transition gate.

\section{Transmon circuit in the charge representation}
\label[appendix]{transmonNoise}
\label[appendix]{sec:decoherence}
\begin{figure*}
    \centering
    \includegraphics[width=\textwidth]{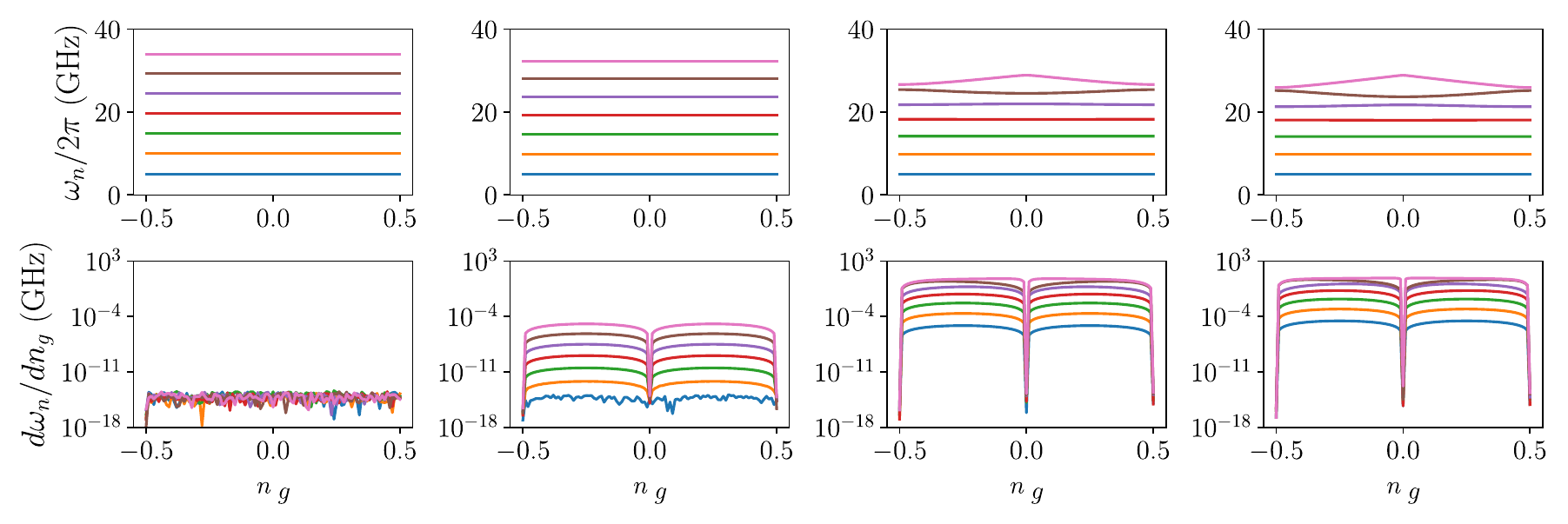}
    \caption{Energy spectrum of the transmon circuit as a function of the dimensionless gate voltage $n_g$ for four different values of $\alpha/(2\pi)=\{-50, -100, -200, -300\}~$(MHz).
    For the numerical simulations, we have fixed $E_J$ for obtaining $\omega_{10}/(2\pi)=5~$(GHz), and as expected, for increasing anharmonicity, the energy spectrum becomes more sensitive to charge fluctuations.
    }
    \label{Fig:SepctrumNg}
\end{figure*}
In this appendix, we will discuss a more general description of the transmon circuit that goes beyond the standard Duffing oscillator~\cite{Blais2021Circuit}. The fundamental aspect of this modelling is the representation of both charge and phase operators in \cref{eq:transmon hamiltonian exact}. In the charge qubit description, the operator $\hat{n}$ describes the excess of Cooper-Pair on the superconducting islands, while the cosine operator $\cos(\hat{\varphi})$ describes the tunnelling between them along the junction. Explicitly, we write:
\begin{eqnarray}
\label{oper_representation}
    \hat{n}&:=&\sum_{n\in \mathbb{Z}}n\ketbra{n}{n},\\
    \exp\big(i\hat{\phi}\big)&:=&\sum_{n\in \mathbb{Z}}\ketbra{n}{n+1}.
\end{eqnarray}
In this representation, the eigenstates of the transmon are obtained by numerically diagonalizing the Hamiltonian to a subspace spanned by a few charge states. This modifies the control operator $\hat{n}$ in such a basis so that it exhibits selection rules different from the bosonic oscillator defining the Duffing oscillator [see \cref{HT}].

The gate voltage $n_g(t)$ responsible for driving the transitions on the transmon circuit is susceptible to fluctuations which could be thermal, due to wiring circuits and quasiparticle tunnelling through the junction, or non-thermal, due to impedance mismatching with the signal generator. Thus, we need to quantify the fluctuation of the energy levels of the transmon by varying $n_g(t)$. \Cref{Fig:SepctrumNg} shows the low-lying energy spectrum as a function of the gate voltage $n_g$. We have selected $E_C$ and $E_J$ such that the $\omega_{10}/(2\pi)=(\omega_1-\omega_0)/(2\pi)=5~$GHz, and we vary the anharmonicity $\alpha=\omega_{21}-2\omega_{10}$ to be in the range $\alpha/(2\pi)=(-50, -300)~$(MHz).

We observe increasing charge dispersion for larger values of $\alpha$. The main reason for the increase of charge dispersion with decreasing $E_{J}$ is the constraint imposed to obtain always the same $\omega_{10}$ frequency; consequently, fewer states are confined in the cosine potential. This feature is more appreciable when we see the variation of the energy spectrum $\partial \omega_{k+1,k}/\partial n_g$ with respect to the gate voltage, where for smaller $\alpha$ the fluctuations are on the order of KHz.

In this scenario, depending on our transmon parameters, we need to carefully select the workable low-lying energy levels for our qudit gates. In our case, we follow a different approach than Ref.~\cite{Wang2025High}; rather than compute the ratio between the deep potential with the energy spacing, we consider the average of the fluctuation over $n_g$. In this work, we set a truncation at $\partial \omega_{N_{\rm{max}}}/\partial n_g \approx10^{-3}$~(GHz) and consider any eigenstates with lower dispersion suitable as a qudit level. This results in the number of levels available in the nonlinear oscillator in \cref{fig:Numberstates}.

This constraint on dispersion also extends to the dephasing time where we have used $1/f$ noise as the most detrimental source of decoherence which can be estimated by the relation~\cite{PhysRevB.72.134519, Koch2007Chargeinsensitive}
\begin{eqnarray}
    \frac{1}{T_{\phi}^{(k)}} = A_{n_{g}} \left| \frac{\partial \omega_{k+1,k}}{\partial n_{g}} \right| \sqrt{2|\ln \left( \omega_{\text{low}} t_{\text{exp}} \right)|},
\end{eqnarray}
where $\omega_{k+1,k}=\omega_{k+1}-\omega_k$  and $A_{n_{g}}=10^{-4}e$ is the noise strength~\cite{PhysRevLett.93.267007, PhysRevB.53.13682, PhysRevB.100.140503} with $e$ being the electron charge. Also, $\omega_{\text{low}}=2\pi/t_{\text{exp}}$ corresponds to the infrared cutoff due to the finite data acquisition time $t_{\text{exp}}=10^{4}~$ns~\cite{Koch2007Chargeinsensitive}

This is illustrated in \cref{fig:T1T2Times}, where one point corresponds to one qudit eigenstate with a specific hardware parameter.
As the anharmonicity decreases, more and more levels with a coherence time longer than $100$~$\mu$s can be included as quantum information registers.

For amplitude damping, we estimate $T_1$ assuming that the main loss mechanism corresponds to capacitive losses. In such a way, Fermi's golden rules give the relation~\cite{smith2020superconducting}
\begin{eqnarray}
\frac{1}{T_{1}^{(k)}} = |\bra{k}\hat{n}\ket{k+1}|^2S(\omega_{k+1,k}),
\end{eqnarray}
where the spectral density for the capacitive losses reads~\cite{PhysRevB.72.134519,PhysRevX.11.011010}
\begin{eqnarray}
    S(\omega_{k+1,k})= \frac{4\hbar E_{C}}{Q_{\text{cap}}(\omega_{k+1,k})} \left[ \frac{\coth \left( \frac{\hbar |\omega_{k+1,k}|}{2 k_B T} \right) }{1+ \exp \left( - \frac{\hbar \omega_{k+1,k}}{k_B T} \right)}\right]~~.
\end{eqnarray}
with $Q_{\text{cap}}(\omega_{k+1,k})=10^{6}(2\pi\times 6~\text{GHz}/|\omega_{k+1,k}|)^{0.7}$~\cite{Braginsky1987,wang2015} the capacitive quality factor per ladder transition.
Also, $k_{B}$ is the Boltzmann constant, and $T=15~$mK is the temperature.
Since this value is not strongly dependent on the levels in our system studied, we do not use it to truncate the qudit level.

For completeness, we plot the $T_1$ for the different energy levels in~\cref{fig:T1T2Times}. We should note that improvement of the coherence times could be possible by implementing different fabrication techniques such as surface error mitigation~\cite{Place2021,tuokkola2024methods}, changing the Niobium with Tantalum as the base superconductor~\cite{Wang2022Practical,Bal2024} or mitigating the micromotion of the circuitry~\cite{kono2024mechanically}, among other techniques. Such shielding on the transmon circuit leads to coherence times nearly in the millisecond scale.

\begin{figure}
    \centering
    \includegraphics[width=0.5\textwidth]{T1T2transmon2.pdf}
    \caption{Coherence times of the transmon circuit as a function of $\alpha$ for different transition frequency $\omega_{k+1,k}$. We set $E_J$ and $E_C$ such that the transition frequency equal to $\omega_{10}/(2\pi)=5~$(GHz). For the amplitude damping, we assume capacitive losses and dephasing correspond to charge fluctuations.}
    \label{fig:T1T2Times}
\end{figure}

\section{Derivation of the Leakage manifold}
\label{sec: Leakage manifold}
Here, we will show that the energy diagram for any qudit gate between the states $(k+1, k)$ is represented as in \cref{fig:qudit energy levels}c. In other words, if we want to implement this single qudit gate, there appears to be a nearly-resonant transition between the states $\ket{k-1}\leftrightarrow\ket{k+2}$. To do so, let us consider the explicit form of the energy of the $k$th energy level after the frame transformation in \cref{quditH} 
\begin{eqnarray}
    \tilde{\Delta}_{k}=\omega_{k}-k(\omega_{k+1}-\omega_{k}-\delta_{d}).
\end{eqnarray}

For the Dufing oscillator model, we know that $\omega_k=\omega_k-\alpha k(k-1)/2$, where $\omega=\sqrt{8E_CE_J}-E_C$ is the transmon frequency, and $\alpha=-E_C$ is the anharmonicity, respectively. Thus, $\Delta_{k-1}=(k-1)(\alpha(k+2)+2\delta_{d})/2$ while $\Delta_{k+2}=(k+2)(\alpha(k-1)+2\delta_{d})/2$. Thus, the detuning between these energy levels is $\delta_{k-1, k+2}=3\delta_{d}$ for all values of $k$, which is zero if the drive is resonant.

However, such a description of the system Hamiltonian is only valid for larger $E_J/E_C$. Thus, for obtaining better estimation of the detuning, we consider the eigenenergies of the transmon obtained by numerical diagonalizing \cref{eq:transmon hamiltonian exact}. \Cref{fig:ga03 sweep}d shows $\delta_{k-1, k+2}$ as a function of the anharmonicity $\alpha$ for several ladder transitions at $n_g=0$; from the figure we appreciate an inverse relation between the degeneracy of the leakage state with the anharmonicity, recovering the previous calculation result when $\alpha=-2\pi\times 50~$(MHz). Moreover, we also see an increase of such discrepancy with the qudit manifold to be addressed, this effect is mainly produced by the sensitivity of the energy spectrum to the charge noise (see \cref{Fig:SepctrumNg}).

\section{Error budget for transmon qudit}
\label{sec:error defintion}

We define the leakage error of a quantum gate as
\begin{equation}
    \mathcal{L}= 1- \mathcal{F}_L =  1 - \frac{\Tr [\hat{\mathcal{U}}_Q \hat{\mathcal{U}}^{\dagger}_Q]}{2}
    ,
\end{equation}
where $\hat{\mathcal{U}}_Q$ denotes the projection of the full unitary evolution onto the two-level target subspace spanned by ${\ket{k}, \ket{k+1}}$.
Correspondingly, leakage to a specific state $\ket{l}$ outside the target subspace is quantified as
\begin{equation}
    \mathcal{L}_{l} = \frac 12 \sum_{j\in\{k,k+1\}} |\mathcal{U}_{l,j}|^2
    \label{eq:leakage}
    ,
\end{equation}
where $\mathcal{U}_{l,j}$ is the matrix element of the full unitary operator.

To capture errors that occur within the computational subspace, such as phase and amplitude errors, we define the subspace error:
\begin{equation}
    \mathcal{E}_Q= 1-\mathcal{F}_Q
    =
    1 - \frac{
    \left|\Tr [\hat{\mathcal{U}}_Q \hat{\mathcal{U}}^{\dagger}_I]\right|^2}{
    2\Tr [\hat{\mathcal{U}}_Q \hat{\mathcal{U}}^{\dagger}_Q]}
    ,
\end{equation}
where $\hat{\mathcal{U}}_I$ is the ideal target unitary (e.g., a $\pi$ or $\pi/2$ gate). This metric captures phase and amplitude errors within the logical subspace.
To distinguish phase error from amplitude error, we optimize the drive amplitude to minimize $\mathcal{E}_Q$, and attribute the residual error to phase mismatch.

Using the above definitions, the total gate fidelity $\mathcal{F}$ in \cref{eq:two level fidelity} can be expressed as:
\begin{equation}
    \mathcal{F}[\hat{\mathcal{U}}_Q]
    =
    \frac 23 \mathcal{F}_Q \mathcal{F}_L + \frac{\mathcal{F}_L}{3}
    .
\end{equation}

\begin{figure*}
    \centering
    \includegraphics[width=\linewidth]{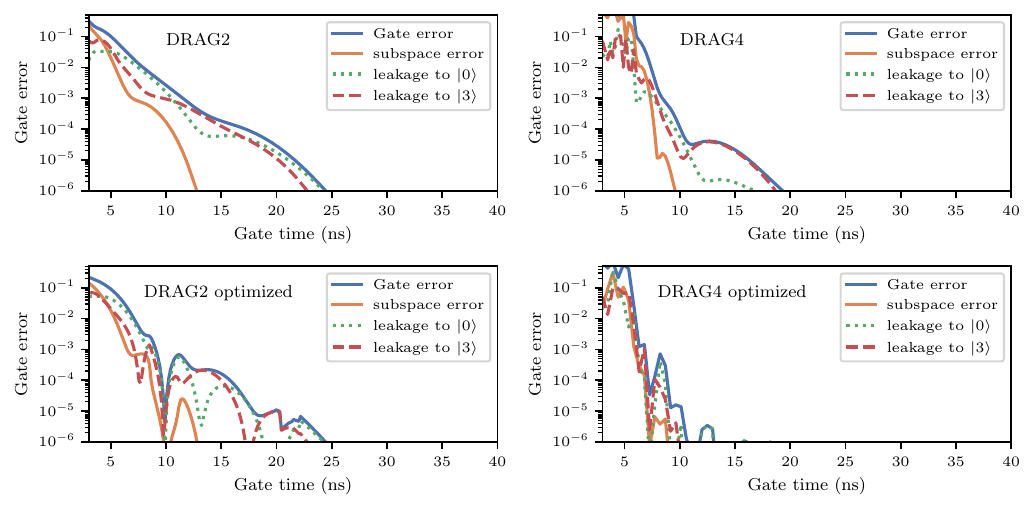}
    \caption{
    Error budget for a $\pi$-rotation between $\ket{1} \leftrightarrow \ket{2}$ in a transmon qudit using different DRAG pulses, with $\alpha/(2\pi) = -200$~MHz. Each panel shows the total gate error alongside contributions from subspace error and leakage to $\ket{0}$ and $\ket{3}$.
    Top row: uncalibrated DRAG2 and DRAG4 pulses; 
    bottom row: same pulses Ansatz after optimizing the DRAG coefficients. 
    }
    \label{fig:error budget DRAG}
\end{figure*}

Following the above defintion, we also studied the error budget for different drive schemes.
\cref{fig:error budget DRAG} complements \cref{fig:error budget}a by providing a detailed breakdown of gate errors under DRAG-based control methods. In the top row, uncalibrated DRAG2 and DRAG4 pulses show that leakage remains the dominant error, which aligns with our observation of the quantum speed limit in~\cref{fig:ladder fidelity vs n}. If additional optimization is performed, as shown in the bottom row, both leakage and subspace errors are further reduced, especially for DRAG4, highlighting the effectiveness of recursive DRAG combined with targeted calibration. Here, we group phase and amplitude errors as "subspace error" since both drive detuning and amplitude are optimized, making it no longer possible to isolate phase error. 
However, we note that full calibration of DRAG4 parameters may increase experimental overhead due to the larger parameter space (see Appendix~\ref{sec: DRAG4 calibration}).

\section{Derivation of recursive DRAG pulse}
\label[appendix]{sec:derivation DRAG}
\subsection{Single-photon correction}
In the following, we show the derivation of the recursive DRAG pulse shape designed to suppress the two single-photon transitions $\ket{k-1}\leftrightarrow\ket{k}$ and $\ket{k+1}\leftrightarrow\ket{k+2}$.
Our general approach is to progressively derive the effective frame and the corresponding drive shapes to minimize the prevalent error.
Throughout the calculation, we keep the perturbative correction up to the second order for all the terms with two exceptions: the matrix entry \((0,3)\), which characterizes a three-photon leakage due to the small energy separation, and the entry \((1,2)\), which describes the pulse amplitude correction.
For those two, we keep the terms up to the third-order correction.

We start with the rotating frame Hamiltonian in \cref{eq:lab 4 level Hamiltonian}
\begin{equation}
    \hat{H}_0 =
 \left(
\begin{array}{cccc}
 -\Delta_{L} & \frac{\lambda_1\bar{\Omega }}{2} & 0 & 0 \\
 \frac{\lambda_1\Omega}{2} & \delta_d  & \frac{\lambda_2 \bar{\Omega }}{2} & 0 \\
 0 & \frac{\lambda_2 \Omega}{2} & 2 \delta_d  & \frac{\lambda_3 \bar{\Omega }}{2} \\
 0 & 0 & \frac{\lambda_3 \Omega}{2} & \Delta_{H} +3 \delta_d \\
\end{array}
\right)
,
\end{equation}
where $\Delta_{H}=\Delta_k+\delta_{k-1, k+2}$ and $\Delta_{L} = -\Delta_k$.
For ease of notation, we use $\lambda_1$, $\lambda_2$, $\lambda_3$ for $\lambda_{k-1}$, $\lambda_k$ and $\lambda_{k+1}$ in this section.
We define the first transition targeting the single-photon leakage error, $\ket{k+1}\leftrightarrow\ket{k+2}$.
For small $\delta_d$, as is typical in the transmon regime, this is the largest leakage source (see \cref{fig:ga03 sweep}).
The frame transformation generator is given by
\begin{equation}
\hat{S}_{0\to 1}=
\left(
\begin{array}{cccc}
 0 & -\frac{\epsilon\lambda_1\bar{\Omega }_1}{2 \Delta_{H}} & 0 & 0 \\
 \frac{\epsilon\lambda_1\Omega_1}{2 \Delta_{H}} & 0 & -\frac{\epsilon\lambda_2 \bar{\Omega }_1}{2 \Delta_{H}} & 0 \\
 0 & \frac{\epsilon\lambda_2 \Omega_1}{2 \Delta_{H}} & 0 & -\frac{\epsilon\lambda_3 \bar{\Omega }_1}{2 \Delta_{H}} \\
 0 & 0 & \frac{\epsilon\lambda_3 \Omega_1}{2 \Delta_{H}} & 0 \\
\end{array}
\right)
,
\end{equation}
where \(\epsilon\) is a small parameter to track the perturbation order.
The denominator $\Delta_{H}$ is chosen such that in the effective frame, the matrix entry $(2,3)$ is zero.
In addition, $\hat{S}_{0\to 1}$ is chosen to be proportional to the control term in $\hat{H}_0$; this is designed in particular such that there is no derivative term $\dot{\Omega}_1$ in $\hat{H}_1$~\cite{Li2024Experimental}.
After substituting the expression \(\Omega = \Omega_1 -\im\frac{\dot{\Omega}_1}{\Delta_{H}}\), we get $\hat{H}_1$ with the off-diagonal term
\begin{align}
&\hat{H}_1 - \hat{H}_{1,\rm{diag}}=
\nonumber
\\
&\left(
\begin{array}{cccc}
 0 & \frac{1}{2} \epsilon  \lambda_{\text{r1}} \bar{\Omega }_1 & \frac{-\Delta_{L}  \epsilon ^2 \lambda _1\lambda_2 \bar{\Omega }_1^2}{8 \Delta_{H}^2} & \epsilon^3\bar{\Omega}_{\text{L03}}^{(1)} \\
 \frac{1}{2} \epsilon  \lambda_{\text{r1}} \Omega_1 & 0 & \frac{ \epsilon  \lambda_2 \left(\bar{\Omega }_1 + \epsilon^2\bar{\Omega}_{\text{c}}^{(1)}\right) }{2} & \frac{\epsilon ^2 \lambda_2 \lambda_3 \bar{\Omega }_1^2}{8 \Delta_{H}} \\
 \frac{-\Delta_{L}  \epsilon ^2 \lambda _1\lambda_2 \Omega_1^2}{8 \Delta_{H}^2} & \frac{ \epsilon  \lambda_2 \left(\Omega_1+\epsilon^2\Omega_{\text{c}}^{(1)}\right)}{2} & 0 & 0 \\
 \epsilon^3\Omega_{\text{L03}}^{(1)} & \frac{\epsilon ^2 \lambda_2 \lambda_3 \Omega_1^2}{8 \Delta_{H}} & 0 & 0 \\
\end{array}
\right),
\end{align}
where $\Omega_{\text{c}}^{(1)}$ and $\Omega_{\text{L03}}^{(1)}$ denote the third order error to the drive amplitude in this frame and the three-photon leakage transition, which we do not explicitly use in the following calculation.
Notice that in the effective frame, we obtain a renormalized leakage rate \(\lambda_{\text{r1}} \Omega_1\) between $\ket{k-1}$ and $\ket{k}$, with \(\lambda_{\text{r1}}=\lambda _1(1-\Delta_{L}/\Delta_{H})\approx 2\lambda _1\) in the limit $\delta_{k-1,k+2}\rightarrow 0$.
This explains why the leakage increases with only one single derivative DRAG correction in \cref{fig:error budget}b.
This prefactor also needs to be taken into consideration when making perturbative assumptions.
The diagonal energy terms are given by 
\begin{align}
&E_{1, \ket{k-1}} = -\Delta_{L}-\frac{\lambda _1^2\Re\left(\Omega \bar{\Omega }_1\right)}{2 \Delta_{H}}+\frac{\lambda _1^2\Delta_{L}  \left| \Omega_1\right| ^2}{4 \Delta_{H}^2},\\
&E_{1, \ket{k}} = \delta_d
+\left(\frac{\lambda _1^2-\lambda_2^2}{2 \Delta_{H}}\right)\Re\left(\Omega \bar{\Omega }_1\right)
- \frac{\lambda _1^2\Delta_{L} \left| \Omega_1\right| ^2}{4 \Delta_{H}^2},
\\
&E_{1, \ket{k+1}} = 2 \delta_d
+\left(\frac{\lambda_2^2-\lambda_3^2  }{2 \Delta_{H}}\right) \Re\left(\Omega \bar{\Omega }_1\right)
+\frac{\lambda_3^2 \left| \Omega_1\right| ^2}{4 \Delta_{H}},
\\
&E_{1, \ket{k+2}} = 3 \delta_d+\Delta_{H}
+\frac{\lambda_3^2 \Re\left(\Omega \bar{\Omega }_1\right)}{2 \Delta_{H}} -\frac{\lambda_3^2 \left| \Omega_1\right| ^2}{4 \Delta_{H}}.
\end{align}

Secondly, we target the single photon leakage between state $\ket{k-1}$ and $
\ket{k}$, with the frame transformation generator
\begin{eqnarray}
\hat{S}_{1\to 2}=
\left(
\begin{array}{cccc}
 0 & -\frac{\epsilon  \lambda_{\text{r1}} \bar{\Omega }_2}{2 \Delta_{L} } & 0 & 0 \\
 \frac{\epsilon  \lambda_{\text{r1}} \Omega_2}{2 \Delta_{L} } & 0 & -\frac{\epsilon  \lambda_2 \bar{\Omega }_2}{2 \Delta_{L} } & 0 \\
 0 & \frac{\epsilon  \lambda_2 \Omega_2}{2 \Delta_{L} } & 0 & 0 \\
 0 & 0 & 0 & 0 \\
\end{array}
\right)
.
\end{eqnarray}
This, together with the substitution \(\Omega_1 = \Omega_2 -\im\frac{\dot{\Omega}_2}{\Delta_{L}}\),  results in the suppression of the transition and gives 
\(\hat{H}_2\)
\begin{widetext}
\begin{align}
&\hat{H}_2 - \hat{H}_{2,\rm{diag}}=
\left(
\begin{array}{cccc}
 0 & 0 & -\frac{\Delta_{L}  \epsilon ^2 \lambda_2 \bar{\Omega }_1^2}{8 \Delta_{H}^2}-\frac{\epsilon ^2 \lambda_2 \lambda_{\text{r1}} \bar{\Omega }_2^2}{8 \Delta_{L} } & \bar{\Omega}_{\text{L03}}^{(1)} \\
 0 & 0 & \frac{ \epsilon  \lambda_2 \left(\bar{\Omega} _2+\epsilon^2\bar{\Omega}_{\text{c}}^{(2)}\right)}{2}  & \frac{\epsilon ^2 \lambda_2 \lambda_3 \bar{\Omega }_1^2}{8 \Delta_{H}} \\
 -\frac{\Delta_{L}  \epsilon ^2 \lambda_2 \Omega_1^2}{8 \Delta_{H}^2}-\frac{\epsilon ^2 \lambda_2 \lambda_{\text{r1}} \Omega_2^2}{8 \Delta_{L} } & \frac{ \epsilon  \lambda_2 \left(\Omega_2+\epsilon^2\Omega_{\text{c}}^{(2)}\right)}{2}  & 0 & 0 \\
 \Omega_{\text{L03}}^{(1)} & \frac{\epsilon ^2 \lambda_2 \lambda_3 \Omega_1^2}{8 \Delta_{H}} & 0 & 0 \\
\end{array}
\right)
.
\label{eq:H after first-order DRAG}
\end{align}
\end{widetext}
In addition to the leakage error, the phase error and the amplitude renormalization also need to be considered to get the desired rotation. The time-dependent phase correction is given by
\begin{align}
    \delta_d &= 
    -\Re\left(\Omega_1 \bar{\Omega }_2\right) \left(\frac{\lambda_2^2 }{\Delta_{L}}-\frac{ \lambda_{\text{r1}}^2}{2 \Delta_{L}}\right)\nonumber\\
    &-\Re\left(\Omega \bar{\Omega }_1\right) \left(-\frac{\lambda_1^2 }{2 \Delta_{H}}+\frac{\lambda_2^2 }{\Delta_{H}}-\frac{\lambda_3^2 }{2 \Delta_{H}}\right)\nonumber\\
    &-\left| \Omega_1\right| {}^2 \left(\frac{\Delta_{L} \lambda_1^2 }{4 \Delta_{H}^2}+\frac{\lambda_3^2 }{4 \Delta_{H}}\right)-\frac{ \left| \Omega_2\right| {}^2 \lambda_{\text{r1}}^2}{4 \Delta_{L}}
    ,
\end{align}
where $\lambda_{r1} = \lambda_1(1 - \Delta_{L}/\Delta_{H})$.
Note that although \(\delta_d(t)\) is promoted to a time-dependent function, this does not affect the derivation of the effective model, as the rotating-frame frequency \(\omega_d\) remains fixed at the idling transition frequency and is independent of the drive.

Apart from that, the correction on the drive shape also slightly affects the rotation angle.
A small correction term needs to be added $\Omega_2 \leftarrow \Omega_2 + \Omega_{\text{amp}}$.
The analytical formula of the amplitude correction is written as

\begin{align}
    \Omega_{\text{amp}}&=\Omega _0 \left| \Omega _1\right| {}^2 \left(\frac{\lambda _1^2}{8 \Delta_{H}^2}+\frac{\lambda _3^2}{8 \Delta_{H}^2}-\frac{\lambda _2^2}{4 \Delta_{H}^2}\right)
    \nonumber\\
    &+\Omega _1 \left| \Omega _2\right| {}^2 \left(\frac{\lambda _{\text{r1}}^2}{8 \Delta_{L}^2}-\frac{\lambda _2^2}{4 \Delta_{L}^2}\right)
    \nonumber\\
    &+\Omega _2\left| \Omega _1\right| {}^2 \left(-\frac{\lambda _1^2}{4 \Delta_{H}^2}-\frac{\lambda _3^2}{4 \Delta_{L} \Delta_{H}}\right)
    \nonumber\\
    &+\Omega _2\Re\left(\Omega _0 \bar{\Omega }_1\right) \left(\frac{\lambda _1^2}{2 \Delta_{L} \Delta_{H}}+\frac{\lambda _3^2}{2 \Delta_{L} \Delta_{H}}-\frac{\lambda _2^2}{\Delta_{L} \Delta_{H}}\right)
    \nonumber\\
    &+\Omega _1^2 \bar{\Omega } \left(\frac{\lambda _1^2}{8 \Delta_{H}^2}+\frac{\lambda _3^2}{8 \Delta_{H}^2}-\frac{\lambda _2^2}{4 \Delta_{H}^2}\right)
    \nonumber\\
    &+\delta _d \left(-\frac{\Omega _1}{\Delta_{H}}-\frac{\Omega _2}{\Delta_{L}}\right)
    \nonumber\\
    &+
    \Omega _1^2 \bar{\Omega }_1\left(-\frac{\Delta_{L} \lambda _1^2}{8 \Delta_{H}^3}-\frac{\lambda _3^2}{8 \Delta_{H}^2}\right)
    \nonumber\\
    &+\Omega _2^2 \bar{\Omega }_1\left(\frac{\lambda _{\text{r1}}^2}{8 \Delta_{L}^2}-\frac{\lambda _2^2}{4 \Delta_{L}^2}\right)
    \nonumber\\
    &-\frac{\lambda _1 \Omega _1^2 \bar{\Omega }_2 \lambda _{\text{r1}}}{8 \Delta_{H}^2}-\frac{\Omega _2^2 \bar{\Omega }_2 \lambda _{\text{r1}}^2}{8 \Delta_{L}^2}
    .
\end{align}
In our investigation, we neglect the time dependence and numerically optimize a fixed correction of the detuning and the amplitude.

\subsection{Two-photon correction}
The first two transitions yield the effective Hamiltonian described in \cref{eq:H after first-order DRAG}, where the desired transition between $\ket{k}\leftrightarrow\ket{k+1}$ is preserved, with a renormalized effective coupling strength.
The diagonalization of the single-photon coupling introduces new two-photon transitions, $\ket{k-1}\leftrightarrow\ket{k+1}$ and $\ket{k}\leftrightarrow\ket{k+2}$, with the coupling strength proportional to $\Omega^2$.
This is also obtained for qubit driving in a nonlinear oscillator, as discussed in~\cite{Motzoi2013Improving}.
For very strong drive amplitude, these two-photon transitions become the dominant source of error once the single-photon transitions are sufficiently suppressed.

For simplicity, we do not repeat the full calculation as in the last subsection but note the following properties.
We can treat $\Omega^2$ as the new coupling $g$, then derive the same expression to suppress the two leakages as in \cref{eq: DRAG2 corrections2} but with the coupling $g$.
Moreover, any perturbative diagonalization of the two-photon transitions will introduce corrections only in the order of $\epsilon^3$ or smaller, which is negligible relative to the truncation order considered.
By substituting $g$ back into $\Omega$, we obtain the expression in \cref{eq: DRAG4 corrections2}.\\

\subsection{Three-photon correction}
In principle, based on the DRAG2 pulse, we can follow a similar strategy and use a recursive DRAG design to suppress the transition error between state $\ket{0}$ and $\ket{3}$:
\begin{equation}
    \Omega_4 = \sqrt[\leftroot{-1}\uproot{2}\scriptstyle 3]{\Omega_5^3-\im \frac{3\Omega_5^2\dot{\Omega}_5}{\delta_{k-1, k+2} }}
    .
\end{equation}
However, due to the small gap between $\ket{k-1}$ and $\ket{k+2}$ in the transmon regime, the imaginary DRAG correction term is much larger and a constant detuning may not suffice to compensate for the phase error.
Nevertheless, even with a not perfectly aligned phase, a $\pi/2$ gate can be implemented with the help of virtual phase gates.

Alternatively, one could explore the direct coupling between the states $\ket{k-1}$ and $\ket{k+2}$ instead of relying on the multi-photon process. However, this would require microwave drive generators with a frequency approximately three times that of the qubit frequency.

\section{Calibration of the DRAG4 pulse}
\label{sec: DRAG4 calibration}

\begin{figure*}
    \centering
    \includegraphics[width=\linewidth]{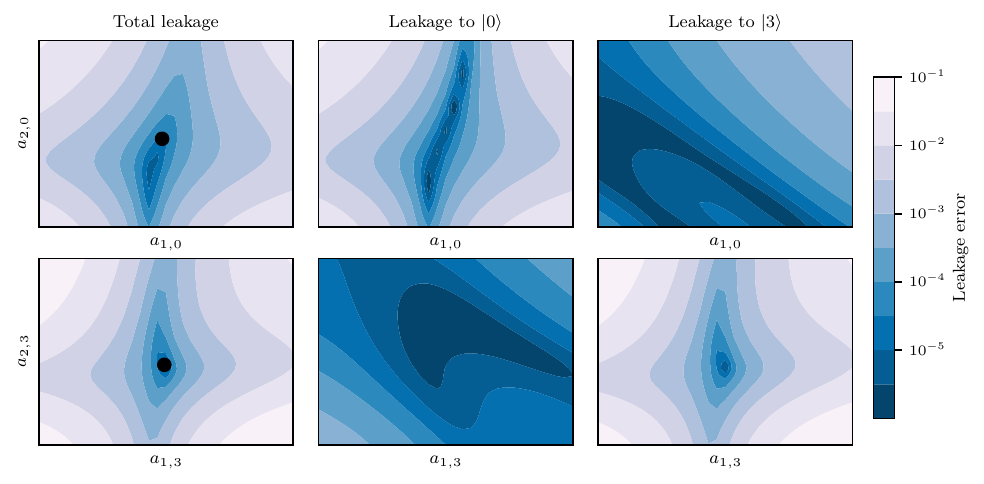}
    \caption{
    Calibration of the DRAG4 pulse. Shown is the leakage error as a function of DRAG coefficients for DRAG4 pulses. A $\ket{1} \leftrightarrow \ket{2}$ $\pi$ gate is calibrated with $\alpha/2\pi = -200$~MHz and a gate duration of 15~ns. The range of all DRAG coefficients is from 0 to 2. In the upper row, the two DRAG coefficients targeting leakage to $\ket{3}$ are fixed at their nominal analytical values (set to one), while the coefficients $a_{1,0}$ and $a_{2,0}$ controlling leakage to $\ket{0}$ are varied. In the lower row, the roles are reversed: $a_{1,0}$ and $a_{2,0}$ are held fixed, and $a_{1,3}$ and $a_{2,3}$ are scanned. The black dot denotes the global minimum of the leakage error across the full parameter space.
    }
    \label{fig:calibration DRAG4}
\end{figure*}
In \cref{fig:calibration DRAG2}, we demonstrated the calibration of DRAG2 pulses. We now extend this analysis to DRAG4, which introduces four tunable parameters: $a_{1,0}$ and $a_{1,3}$ for suppressing single-photon leakage to $\ket{0}$ and $\ket{3}$, and $a_{2,0}$ and $a_{2,3}$ for mitigating two-photon leakage to the same states.

\Cref{fig:calibration DRAG4} illustrates the leakage error landscape by scanning pairs of DRAG4 coefficients while holding the remaining two fixed at their nominal analytical values. As expected, $a_{1,0}$ and $a_{2,0}$ primarily influence leakage to $\ket{0}$, while $a_{1,3}$ and $a_{2,3}$ control leakage to $\ket{3}$. However, in contrast to the DRAG2 case, the two coefficients associated with the same leakage level (e.g., $a_{1,0}$ and $a_{2,0}$) cannot be calibrated independently, since both require measurement of the same leakage population. This complicates the calibration process.

The global minimum of the leakage error in each parameter space is indicated with a black dot in the left two plots. For $\ket{0}$-leakage, the optimal point deviates significantly from the nominal values ($a_{n,l} = 1$), indicating that independent calibration of each coefficient may result in suboptimal performance or require a more extensive search. Nonetheless, hill-climbing optimization such as Nelder-Mead would be a good option requiring few iterations. A potentially more efficient calibration strategy might involve fixing part of the coefficients, such as the DRAG2 terms $a_{1,0}$ and $a_{1,3}$, and optimizing the remaining pair. We leave further exploration of such hybrid calibration schemes for future work.

In summary, while the four DRAG4 coefficients offer additional degrees of freedom that can significantly improve gate fidelity, they also increase the complexity of calibration in practice.

\begin{acknowledgments}
The authors would like to thank José Jesus and Shahrukh Chishti for insightful discussion.
This work was funded by the Federal Ministry of Education and
Research (BMBF) within the framework programme "Quantum technologies – from
basic research to market" (Project QSolid, Grant No.~13N16149),
by the Deutsche Forschungsgemeinschaft (DFG, German Research Foundation) under Germany's Excellence Strategy – Cluster of Excellence Matter and Light for Quantum Computing (ML4Q) EXC 2004/1 – 390534769,
by HORIZON-CL4-2022-QUANTUM-01-SGA Project under Grant 101113946 OpenSuperQPlus100 and the European Union’s Horizon Programme (HORIZON-CL4-2021-DIGITALEMERGING-02-10) Grant Agreement 101080085 QCFD.
AL acknowledges the support of the Natural Sciences and Engineering Research Council of Canada (NSERC), through the Alliance International Catalyst Quantum Grant ALLRP 580928 - 22.
The authors gratefully acknowledge the Gauss Centre for Supercomputing e.V. (www.gauss-centre.eu) for funding this project by providing computing time through the John von Neumann Institute for Computing (NIC) on the GCS Supercomputer JUWELS at Jülich Supercomputing Centre (JSC).
\end{acknowledgments}

\end{document}